\documentclass[fleqn,usenatbib,useAMS]{mnras}

\usepackage{graphicx}	
\usepackage{amsmath}	
\usepackage{amssymb}	
\usepackage{multicol}        
\usepackage{pdflscape}	
\usepackage{csvsimple}
\usepackage{rotating}
\usepackage{pdfpages}
\usepackage{caption}
\usepackage{url}
\usepackage{threeparttable}
\usepackage{tabularx}
\usepackage{physics}
\usepackage{siunitx}
\usepackage{todonotes}
 
\DeclareSIUnit{\erg}{erg}
\DeclareSIUnit{\parsec}{pc}
\DeclareSIUnit{\yr}{yr}
\DeclareSIUnit{\seconds}{s}
\graphicspath{{figures/}}

\usepackage[T1]{fontenc}
\usepackage{ae,aecompl}

\usepackage{newtxtext,newtxmath}

\author[Grunthal et al.]{ 
K. Grunthal$^1$\thanks{E-mail: s6kagrun@uni-bonn.de},
M.~Kramer$^{1,2}$\thanks{E-mail: mkramer@mpifr-bonn.mpg.de}
G.~Desvignes$^{3,1}$
\\
$^1$ Max-Planck-Institut f\"{u}r Radioastronomie, Auf dem H\"{u}gel 69, D-53121 Bonn, Germany\\
$^2$ Jodrell Bank Centre for Astrophysics, University of Manchester, M13 9PL, UK\\
$^3$ LESIA, Observatoire de Paris, Université PSL, CNRS, Sorbonne Université, Université de Paris, 5 Place Jules Janssen, 92195, Meudon, France
}

\title[DNS merger rates]
{Revisiting the Galactic Double Neutron Star merger and LIGO detection rates}

\date{Last updated; in original form}

\pubyear{2021}

\begin{document}
\label{firstpage}
\pagerange{\pageref{firstpage}--\pageref{lastpage}}
\maketitle

\begin{abstract}
We revisit the merger rate for Galactic double neutron star (DNS) systems in
light of recent observational insight into the longitudinal and latitudinal
beam shape of the relativistic DNS PSR J1906$+$0746. Due to its young age and its relativistic orbit,
the pulsar contributes significantly to the estimate of the joint Galactic 
merger rate. We follow previous analyses by modelling the underlying pulsar
population of nine merging DNS systems and study the impact and resulting 
uncertainties when replacing simplifying assumptions made in the past with
actual knowledge of the beam shape, its extent and the viewing geometry. 
We find that the individual contribution
of PSR J1906$+$0746 increases to $\mathcal{R} = 6^{+28}_{-5} \si{\per\mega\yr}$ although the values is still consistent with previous estimates  given the
uncertainties. We also compute contributions
to the merger rates from the other DNS systems by applying a generic beam shape
derived from that of PSR~J1906+0746, evaluating the impact of 
previous assumptions.
We derive a joint Galactic DNS merger rate of
$\mathcal{R}^{\mathrm{gen}}_{\mathrm{MW}} = 32^{+19}_{-9}\ \si{\per\mega\yr}$,
leading to a LIGO detection rate of
$\mathcal{R}^{\mathrm{gen}}_{\mathrm{LIGO}} = 3.5^{+2.1}_{-1.0}\ \si{\per\yr}$
(90\% conf.\ limit), considering the upcoming O3 sensitivity of LIGO.
As these values are in good agreement with previous estimates, we conclude that the method of estimating the DNS merger and LIGO detection
rates via the study of the radio pulsar DNS population is less prone
to systematic uncertainties than previously thought. 
\end{abstract}

\begin{keywords}
pulsars:general, stars:neutron
\end{keywords}



\section{Introduction}

Double neutron star (DNS) systems are of major interest for a number
of reasons. Firstly, their formation invokes a number of questions
about the progenitor systems, the birth of neutron stars themselves
via different possible mechanisms, and their size and origin of
resulting properties like mass, spin or orbital configuration
\citep{tauris17}. Secondly, if (at least) one of the neutron stars is
visible as a radio pulsar, DNS systems can act as tools for precision
tests of strong-field gravity \cite[e.g.~][]{ksm+06}. Indeed, the
first DNS system known (PSR B1913+16), discovered in 1974 and now
widely known as the Hulse-Taylor Pulsar \citep{hulse_taylor}, provided
the first evidence for gravitational waves (GW) \citep{tw82}. Such
DNSs visible as binary pulsars are rare, and only about 20 DNS systems
have been detected in radio surveys to date
(e.g.~\citealt{tauris17,hfp+21}). This is both due to their relative
intrinsic rareness caused by the conditions during the formation
process but also due to the difficulties in detecting the pulsar in
fast binary orbits (e.g.~\citealt{tauris17}).

The third reason, why these systems are of interest, is given by the
fact that for compact DNS, the orbital decay of these systems
inevitably leads to the merger of the two neutron stars, an event that
creates a copious amount of GWs during the merger which can be picked
up by terrestrial GW detectors. Indeed, following the discovery of the
Hulse-Taylor pulsar, expected to merge with 300 Myr, it was
anticipated, that the first GW event seen by ground-based
observatories, such as LIGO (Laser Interferometer Gravitational-Wave
Observatory) or Virgo, would be a merger of two NSs
\citep{Barish:1999sy}.  Consequently, estimates of the DNS merger rate
have been made frequently, based on population synthesis calculation
or computations informed by the population of compact DNS observed as
radio pulsars.

Interestingly, the first detected GW merger signal originated from a
binary black hole, while GWs coming from an DNS merger were not
detected until August 2017. This latter event named GW170817
\citep{PhysRevLett.119.161101} revealed an electromagnetic counterpart
which was followed up across the electromagnetic spectrum, truly
starting the era of multi-messenger astronomy
\citep{Abbott_2017}. Since then, however, only one additional
DNS-merger event, GW190425, has been detected
\citep{Abbott_2020}. Hence, improving our knowledge of the DNS merger
rate is still of great scientific interest. On one hand, it will help
predicting the actual event rate, preparing also electromagnetic
follow-up observations. On the other hand, once the actual event rate
has been measured more precisely in upcoming detector runs, we can in
turn use the comparison with theoretical predictions based on the
current understanding of DNS populations in order to re-calibrate our
knowledge about the formation and evolution of Galactic DNS systems.

The detection rate for ground-based GW detectors can be estimated
based on the assumption that the DNS merger rate in the Milky Way can
be extrapolated to external galaxies in the GW detector range.
Essentially, one can obtain the expected Galactic DNS merger rate via
two routes. One is the computation by first principles, using ab
initio population synthesis computations
(e.g.~\citealt{bkk+08}). Another way is to infer the merger rate from
the known Galactic population of DNS systems, which are observed as
binary radio pulsars. However, discovering the relevant DNS systems,
i.e.~those that merge due to the emission of GWs within a Hubble time,
is typically subject to a number of selection effects \citep{lk05}.
Therefore, it is necessary to use methods which take these into
account.

\cite{Kim_2003} developed a method to derive a merger rate probability
distribution using the known DNS population and Bayesian
statistics. The combination of all individual DNS contribution then
yields the merger rate of the Milky Way $\mathcal{R}_\mathrm{MW}$. By
extrapolating that rate to the observable volume of the LIGO/Virgo
system, the number of DNS mergers that the observatories will detect,
can be predicted. The method was also recently applied by
\cite{Pol_2019,Pol_2020}. In the latter update, the authors presented
a Milky Way merger rate of
$\mathcal{R}^{\mathrm{Pol}}_{\mathrm{MW}} = 37 ({+24},{-11})\
\si{\per\mega\yr}$ and an inferred LIGO detection rate of
$\mathcal{R}^{\mathrm{Pol}}_{\mathrm{LIGO}} = 4.2 ({+2.6}, {-1.3})\
\si{\per\yr}$ (90\% confidence limit, c.l.). The extrapolation from
$\mathcal{R}^{\mathrm{Pol}}_{\mathrm{MW}}$ to
$\mathcal{R}^{\mathrm{Pol}}_{\mathrm{LIGO}}$ was done for the LIGO O3
range distance of $\SI{130}{\mega\parsec}$.

There are three factors that determine the rate constraints for pulsar
binaries, namely the number of suitable binary pulsars, their
effective lifetime and their beaming fraction, i.e.~the fraction of
the sky potentially illuminated by a given pulsar. The number of
pulsars can be estimated via survey simulations as explained further
below, which takes the sensitivity of past and current observations
(including selection effects due to binary motion), as well as the
luminosity distribution of the sources into account.

Following~\cite{O_Shaughnessy_2010} and \cite{Pol_2019}, the lifetime
of a pulsar binary can be estimated from the sum of the current age of
the system and the remaining detectable lifetime. The age can be
estimated from the current spin-properties. For a non-recycled pulsar,
one uses the characteristic age (e.g. \cite{lk05}). For a recycled
pulsar, one compares the rotation period and spin-down rate with their
combination given by the so-called ``spin-up line''.  The remaining
lifetime is either the time that the pulsar needs to spin-down until
it crosses the pulsar death line (see e.g.~\cite{lk05}) or until it
merges with its companion, whatever is smaller.

The beaming fraction takes into account that radio surveys only detect
those systems, where the pulsar beam is directed towards Earth,
whereas GW detectors are unaffected by such selection effects.
Following \cite{Kim_2003} or \cite{O_Shaughnessy_2010}, one can
introduce the {\em pulsar beaming correction factor}, $f_\mathrm{b}$,
as the inverse of the fraction of the sky that is illuminated by a
given pulsar's beam (accounting for both magnetic poles). The beaming
fraction depends on the geometry of the pulsar, i.e.~the magnetic
inclination angle between the spin and magnetic axes, $\alpha$, and
the angular size of the emission beam, $\rho$.  These two quantities
also affect the pulse width, $W$, of the observed pulse profile, which
obviously is only a one-dimensional cut through an emission beam with
both longitudinal and latitudinal dimensions. In order to estimate the
overall illuminated fraction of the sky, one usually makes the
assumption that the inferred longitudinal dimension also applies to
the latitudinal direction, i.e.~a circular beam
shape. 

But even the inference of the longitudinal dimension, based on the pulse width, is often
problematic as it requires, apart from $\alpha$ and the assumption that emission completely
fills the pusar beam area, also knowledge of the {\em impact angle}, $\beta$, i.e.~the smallest angular separation of our line-of-sight with the magnetic axis. In this case, 
\begin{equation}
\label{equ:width}
\cos\rho = \cos\alpha\cos(\alpha+\beta)+\sin\alpha\sin(\alpha
+\beta)\cos\left(\frac{W}{2}\right)
\end{equation}
\cite{ggr84}. In principle, it is possible to determine $\alpha$ and $\beta$ from polarisation information by
applying the so-called ``Rotating Vector Model'' (RVM) \citep{rc69}. However, uncertainties due to co-variances in
the parameters can be large (e.g.~\citealt{lk05}), while it is not clear how well the RVM is applicable
to recycled pulsars (see e.g.~\citealt{ksv+21}).

In cases, where the geometry cannot be determined, one can
resort to apply knowledge inferred for population properties to estimate $\rho$.
Based on the analysis of non-recycled pulsars, a 
variety of authors (e.g.~\cite{ran93,gks93,kwj+94,gl98}) has determined that the
value of $\rho$ scales with the pulse period as $\rho = \mathbb{A} \times P^{-0.5}$, whereas $\mathbb{A}$ ranges from
4.9 to 6.5 deg s$^{0.5}$, depending on the study and the intensity level that the width, $W$, was measured for
(see the discussion by \citealt{vbv+19}). This scaling relationship correctly reflects
the period-dependency as expected for a dipolar field line structure. Nevertheless, 
empirically the relationship breaks 
down for recycled pulsars (at about $P\sim10 - 30$ ms), 
which typically appear to show a smaller beam size than 
expected from period scaling \citep{kxl+98}.

\cite{O_Shaughnessy_2010} provided a careful analysis, combining knowledge from
the known population of recycled and non-recycled pulsars and all possible 
information for the visible pulsars in DNS systems known at the time, in order to derive
an effective beaming correction factor, $f_{\rm b,eff}$.
In a recent variation of this work, \cite{Pol_2019} also adopted overall the approach
by \cite{O_Shaughnessy_2010} (and earlier by \cite{Kim_2003}), to assume a fixed  $f_{\rm b,eff}$
value (set to 4.6), while adopting the actually inferred beaming correction factor $f_{\rm b}$
for those three DNS with a determined geometry, i.e.~PSRs J0737$-$3039A, B1534$+$12 and B1913$+$16.
Given the crucial nature of the beaming correction factor, it is desirable to estimate the impact of the simplifying assumptions made.

Recently, long-term observations of the relativistic binary pulsar PSR
J1906$+$0746 allowed to map the emission beam of the visible, non
recycled pulsar of this DNS also in latitudinal direction. This is
possible due to the effects of relativistic spin-precession, which
slowly changes the impact angle and hence the location of the
one-dimensional cut through the beam \citep{Desvignes1013}.  Spanning
observations of more than 10 years, \citet{Desvignes1013} were able to
determine the beam shape, even obtaining information from both
magnetic poles (via the ``main pulse'' and ``interpulse'' emission
separated by half a period) that can be combined to estimate for the
first time reliable the previously unknown $f_\mathrm{b}$ and the
exact viewing geometry of the pulsar (see Section~\ref{sec:1906}).

These measurements show additionally that the beam intensity profile
can neither be modeled by a box-shaped function with the opening angle
as the radius, nor that the main pulse and the interpulse have the
same intensity profile and intensity maxima (for a given magnetic
latitude), as it is usually assumed.
As PSR J1906$+$0746 significantly contributes to
$\mathcal{R}^{\mathrm{Pol}}_{\mathrm{MW}}$
\citep{lsf+06,O_Shaughnessy_2010}, it is therefore an ideal test case
to study the impact of making particular assumptions on the robustness
of the derived DNS merger rate estimates. One example that we can
study in detail is indeed the importance of the applied certain
beaming correction factors, such as assuming a uniform value rather
than using one that is better reflecting the intrinsic properties of a
given pulsar.  Such a study does not only allow us to evaluate the
robustness of the simulations and to better understand the
uncertainties, but also provides us with an updated merger rate.

The plan for the paper is as follows. Firstly, we briefly summarise
the information about PSR J1906$+$0746, which plays a central role in
our study by providing direct and accurate measurements of the extent
of an emission beam and hence the beaming fraction. We then summarise
the statistical approach adopted in this work, following
\cite{Kim_2003} and \cite{Pol_2019}. This is followed by a description
how we use the new information on PSR J1906$+$0746 to improve on the
previous studies, which we extend then also to other DNS systems,
before we discuss the results and draw conclusions.

\section{PSR J1906$+$0746}
\label{sec:1906}

Pulsar J1906$+$0746 was discovered in 2006  \citep{lsf+06}. The 144-ms pulsar with a small characteristic age of only 113\,kyr has an unseen companion in an eccentric ($e=0.085$) 4-hr 
orbit. Even though the spin properties of the pulsar suggested that it is young and unrecycled,
the exact nature of the binary companion was initially unclear. The matter was settled,
when  \citet{vks+15} presented an improved timing solution including the measurement of three post-Keplerian orbital parameters (e.g.~\cite{lk05}), resulting in mass measurements for the 
pulsar of $1.291\pm0.01 M_\odot$, and for the companion of $1.32\pm0.01 M_\odot$, respectively. 
Combined with the orbital parameters, the masses identify the companion as another, recycled but unseen NS, while making PSR J1906$+$0746 the second-born NS in the system. 

After the discovery, the pulsar was identified also in archival data, revealing that by the time of the discovery, emission from the second magnetic pole had became visible as an ``interpulse'', which was missing earlier \citep{lsf+06}. The origin of this change in the observed emission is caused by a change in viewing geometry due to relativistic spin precession that was studied in great detail by \cite{Desvignes1013}. Desvignes et al. were not only able to provide the best test of relativistic
spin precession as predicted by General Relativity thus far, but they were also able to determine
the viewing geometry accurately. They found the inclination angle of the magnetic axis related to the ``main pulse'' to be $\alpha=99.4\pm0.2$ deg,  and a large tilt-angle of the spin axis of the pulsar to the total angular momentum vector of $\delta = 104\pm9$ deg. The observed pulsed emission evolves with time, with the main pulse now having almost disappeared from sight, while the interpulse has remained to be observable for the moment. The slowly changing line-of-sight through the emission beams provides a ``tomographic'' view of the emission regions as it is projected on the plane of the sky. In contrast to existing phenomenological models (see \cite{lk05} for an overivew), the emission is not
symmetric about the magnetic axis, neither in structure nor in extent. While the longitudinal extent of the emission beams varies between 10 and 20 deg, the latitudinal extent is observed to be about 20 deg. Most importantly, however, the luminosity across the beam is highly non-uniform, decreasing
noticeably before the beam ``edge'' is reached. This is in sharp contrast to the simple conal beam model with uniform luminosity that has been applied so far in studies to infer the DNS merger rate.

Addressing the deviation from the simple model used so far is, actually, of particular importance for PSR J1906$+$0746. With its young age, scale factors inferred to 
estimate the number of similar pulsars existing in the Milky Way, are 
unusually large, making the contribution of PSR J1906$+$0746 to the overall estimated merger 
rate comparable to that of the Double Pulsar\footnote{The Double Pulsar has a merger
timescale of only 85 Myr compared to 300 Myr for J1906$+$0746.}, 
as already pointed out by \cite{O_Shaughnessy_2010}. It is therefore especially important to revisit the previously inferred 
merger rate in light of the new recent information provided by \cite{Desvignes1013}.

\section{Calculating DNS merger rates from simulations with the PSRPOPPY package}

Since all known pulsars only represent a small fraction of the total pulsar population in the Milky Way Galaxy, it is necessary to use models and simulations to account for the various selection effects
described earlier in order to estimate the underlying  population. We follow previous works, in particular that of \cite{Pol_2019} and use the simulation code \texttt{PsrPopPy2}\footnote{https://github.com/devanshkv/PsrPopPy2} as well as the separate, available analysis 
code used by \citet{Pol_2019}\footnote{https://github.com/NihanPol/2018-DNS-merger-rate}. While we modify
the code as described later to enable our updated analysis, we maintain the basic approach
applied by \cite{Pol_2019} to obtain an estimate for the size of the Galactic population of pulsars, $N_{\mathrm{tot}}$ and to subsequently infer the rate of DNS mergers to be detected by LIGO. We
describe this method in the following to a detail that is sufficient to motivate and 
clarify our modifications that we explain in Section \ref{sec:enhancements}.

\subsection{Statistical analysis towards the LIGO NS-NS merger detection rate}
\label{sec:poletal}
  
In order to estimate the total Galactic DNS population, the basic idea presented
by \cite{Kim_2003} and also applied by \cite{Pol_2019} is simple. It is
assumed that each currently detected DNS system originates from a separate part of a total DNS
population that represents the individual {\em current} properties of the system and its visible pulsar. 
The systems are therefore treated independently and not, for example, as an evolved version of another observed DNS system. While this {\em snapshot analysis} circumvents the need for sophisticated
binary evolution codes (which may bring uncertainties by themselves), the method still needs to
reflect the selection effects, such as probability of the pulsar beam pointing towards Earth.
Thus, the population size behind any detected DNS pulsar must be of a particular size, so that statistically exactly one pulsar is seen with the sensitivities of the conducted radio surveys.
The statistical considerations leading to the population size that returns one detected pulsar are derived by \citet{Kim_2003}. 

As $N_{\mathrm{tot}} \gg N_{\mathrm{obs}}$, it is expected that the probability of observing $N_{\mathrm{obs}}$ pulsars out of an $N_{\mathrm{tot}}$-sized population follows the Poisson distribution 
\begin{equation}
\label{eq:poisson}
P_{\mathrm{Pois}}(N_{\mathrm{obs}}; \lambda) = \frac{\lambda^{N_{\mathrm{obs}}}e^{-\lambda}}{N_{\mathrm{obs}}!}
\end{equation}
where, by definition, $\lambda \coloneqq \langle N_{\mathrm{obs}}\rangle$. By varying $N_{\mathrm{tot}}$, \cite{Kim_2003} also found that
\begin{equation}
\label{eq:gamma}
\lambda = \gamma N_{\mathrm{tot}}
\end{equation}
with the constant $\gamma$ depending on the properties of the pulsar population. By knowing $\gamma$ one can derive $N_{\mathrm{tot}}$ from $N_\mathrm{obs}$.

The final goal is estimating the Galactic DNS merger rate from simulated populations, where in the first step, the likelihood function $P(\lambda|DX)$ is derived, i.e.\ the probability to obtain a model hypothesis, here the expectation value $\lambda$, given a data set of pulsars $D$ and the model priors $X$ (physical parameters of each pulsar, e.g.\ luminosity, pulse period etc.). Since the only probability known so far is the probability to observe a sample of pulsars $D$ with given $\lambda$ and $X$, i.e.
\begin{equation}
P(D|\lambda X) = P_{\mathrm{Pois}}\left(1; \lambda(N_{\mathrm{tot}}), X\right) = \lambda(N_{\mathrm{tot}}) e^{-\lambda(N_{\mathrm{tot}})},
\end{equation}
the Bayes' theorem is applied to rewrite $P(\lambda|DX)$ in terms of $P(D|\lambda X)$ as
\begin{equation}
\centering
P(\lambda|DX) = P(\lambda|X) \cdot\frac{P(D|\lambda X)}{P(D|X)}
\end{equation}
Following the arguments given by \cite{Kim_2003}, the likelihood function of $\lambda$ then is given as
\begin{equation}
\centering
P(\lambda) = P(\lambda|DX) = P_{\mathrm{Pois}}\left(1;\lambda(N_{\mathrm{tot}}), X\right) = \lambda(N_{\mathrm{tot}}) e^{-\lambda(N_{\mathrm{tot}})}.
\end{equation}
From there, the probability distribution for $N_{\mathrm{tot}}$ can easily be obtained as
\begin{equation}
\label{eq:ntotrate}
P(N_{\mathrm{tot}}) = P(\lambda) \left|\frac{\mathrm{d}\lambda}{\mathrm{d}N_{\mathrm{tot}}}\right| = \gamma^2N_{\mathrm{tot}} \cdot e^{-\gamma N_{\mathrm{tot}}}
\end{equation}
where Equation (\ref{eq:gamma}) is used.

Furthermore, the merger rate $\mathcal{R}$ can be calculated via
\begin{equation}
\centering
\label{eq:ratewithfb}
\mathcal{R} = \frac{N_{\mathrm{tot}}}{\tau_{\mathrm{life}}}f_\mathrm{b}
\end{equation}
for a given population size $N_{\mathrm{tot}}$, with the beaming correction factor $f_\mathrm{b}$ and their lifetime $\tau_{\mathrm{life}}$. Using this equation, the desired merger rate probability is given as
\begin{equation}
\label{eq:mergerrate}
P(\mathcal{R}) =\left(\frac{\gamma\tau_{\mathrm{life}}}{f_\mathrm{b}}\right)^2\cdot \mathcal{R}\cdot e^{-(\gamma\tau_{\mathrm{life}}/f_\mathrm{b})\mathcal{R}}
\end{equation}

In order to determine an estimate of the number of DNS merger events to be 
detected by the LIGO/Virgo network, at first the Milky Way merger rate $\mathcal{R}_{\mathrm{MW}}$ is determined via fitting Eqn.\ \ref{eq:mergerrate} to a given data set. Then this rate is extrapolated to the observable volume of LIGO.
To this end, one must estimate the formation rate of DNS systems in other galaxies. 

In the nearby Universe, the formation rate of binary compact objects is expected to be proportional to the star formation rate, where a measure for the same is the $B$-band luminosity of the given galaxy\footnote{The $B$-band luminosity is the blue luminosity $L_B$ extracted from the galaxy's absolute blue magnitude $M_B$.} \citep{Kopparapu_2008,Phinney_1991}. Inside a sphere with radius $r$, the DNS merger rate follows
\begin{equation}
\mathcal{R}_{\mathrm{LIGO}} = \mathcal{R}_{\mathrm{MW}} \frac{L_{\mathrm{total}}(r)}{L_{\mathrm{MW}}}
\end{equation}
with the total blue luminosity $L_{\mathrm{total}}(r)$ within the distance $r$ and the $B$-band luminosity of the Milky Way $L_{\mathrm{MW}} = 1.7 \times 10^{10} L_{B,\odot}$ denoted in terms of the solar $B$-band luminosity $L_{B,\odot} = 2.16 \times 10^{33}\si{\erg\per\second}$ \citep{Kopparapu_2008}. The actual (Advanced) LIGO range\footnote{Radius of a Euclidean sphere containing the same volume as the true redshifted volume \citep{chen2017distance}.} in run O3 after the last upgrade is estimated to be $D_r^{LIGO} = \SI{130}{\mega\parsec}$ \citep{PhysRevLett.119.161101,Abbott2018}. Following the arguments laid out by \cite{Kopparapu_2008}, the rate of DNS mergers seen by LIGO is given by (cf.\ Eqn.~(19) in \cite{Kopparapu_2008})
\begin{equation}
\label{eq:ligorate}
\mathcal{R}_{\mathrm{LIGO}} = 7.4 \cdot 10^{-3} \left(\frac{\mathcal{R}_{\mathrm{MW}}/L_{\mathrm{MW}}}{(10^{10}L_{B,\odot})^{-1} \si{\mega\parsec}}\right) \times  \left(\frac{D_h^{\mathrm{LIGO}}}{\SI{100}{\mega\parsec}}\right) \si{\per\yr}
\end{equation}
where $D_h^{LIGO} = 2.28 \cdot D_r^{LIGO}$  is the horizon distance\footnote{The
horizon distance is the farthest luminosity distance a source could be detected above 
a certain threshold (\cite{chen2017distance}).} of LIGO in run O3 \citep{chen2017distance,Chernoff_1993}. In the derivation of the formula, the sensitivity decrease of LIGO depending on the orientation of the GW source with respect to the ground-based detector is already taken into account, as pointed out recently by \cite{Pol_2020}. Therefore this correction to Eqn.~(15) in \cite{Pol_2019} is also applied here.

\subsection{PsrPopPy2}
\label{sec:psrpoppy}

In order to infer the underlying Galactic radio pulsar population in the presence of selection effects, we follow the example of \cite{Pol_2019} and deploy the package {\tt PsrPopPy2}.
The package provides a wide range of tools for the simulation and modelling of the pulsar 
populations and their evolution as described in  \cite{psrpoppy}.
The simulations are done within {\tt PsrPopPy2} creating a pulsar population with given pulsar parameters 
and then analysing their detection 
during one or many chosen model surveys that reflect the real-life observing campaigns that
led to the observed population.
In the ``snapshot method'', the aim is to determine $N_{\mathrm{obs}}$, the number of pulsars detected by a survey, depending on the total number of pulsars in the population $N_{\mathrm{tot}}$. 
The analysis is based on a Monte-Carlo-simulation: A number $N_{\mathrm{tot}}$ of pulsars in the Galaxy is created, where each pulsar is given random parameters in luminosity and physical properties, 
drawn from a number of models distributions each (see \citealt{psrpoppy}).
After that, the population is run through a simulated radio telescope survey. Based on the specific survey parameters encoded in the {\tt PsrPopPy2} framework, it is checked how many pulsars could have been detected on Earth, which is then used as follows.

The goal is to obtain $\gamma$ of Eqn.~\ref{eq:gamma} for each kind of pulsar. 
Populations with numbers of pulsars, $N_{\rm tot}$, varying between $10$ and a few thousand are 
created and subjected to the simulated surveys, saving the number of observed pulsars,
$ N_{\mathrm{obs}}$. 
That yields a dataset, to which a Poisson curve  (cf.\ Eq.\ \ref{eq:poisson}) is fitted to obtain $\lambda$  for each $N_{\mathrm{tot}}$. 
With this information, $\gamma$ is calculated for each pulsar by a linear fit to all pairs of ($N_{\mathrm{tot}}$, $\lambda$), following equation (\ref{eq:gamma}).
The size of the expected galactic DNS (pulsar) population $N_{\mathrm{real}}$, 
i.e.\ the population size $N_{\mathrm{tot}}$, that yields one detection, is given by 
rearranging equation \ref{eq:gamma} and inserting $\lambda = \langle N_{\mathrm{obs}}\rangle = 1$.
Therefore $N_{\mathrm{real}}$ is given by $1/\gamma$.
Then $N_{\mathrm{real}}$ and $\gamma$ are used to calculate the probability density function (PDF) for the population size and the merger rate following equations (\ref{eq:ntotrate}) and (\ref{eq:mergerrate}). \\

\section{Utilizing information on the beam shape of PSR J1906$+$0746}
\label{sec:enhancements}

\subsection{Beam tomography}

As relativistic spin-precession causes the viewing geometry of PSR~J1906$+$0746 to change,
our line-of-sight (LOS)
cuts through the emission beam in different ways. This makes it possible to use the observed 
pulse profiles to construct a latitudinal cross section of the pulsar beam.
In the case of PSR J1906$+$0746, the geometry is such that the angle between the
magnetic and spin axes is measured to be $99.4\pm 0.2$ deg. As a result, for 
a period of about 20 years,
both magnetic poles were visible to an observer on Earth as two pulses separated 
from each other by half a spin period. Due to an inhomogeneity in the pulsar beams, the 
pulses observed from each pole 
are not of equal or uniform intensity but in fact differ and change with time 
and relative intensity.
The pole, the emission of which was stronger at discovery, is producing the ''main pulse'' (MP),
whereas the other pole causes the observed ''interpulse'' (IP). As the line-of-sight moved out
of the ''main'' beam, its observed intensity faded progressively, making the second
pulse, the interpulse, the currently stronger one observed.

\begin{figure}
	\centering
	\includegraphics[width=\linewidth]{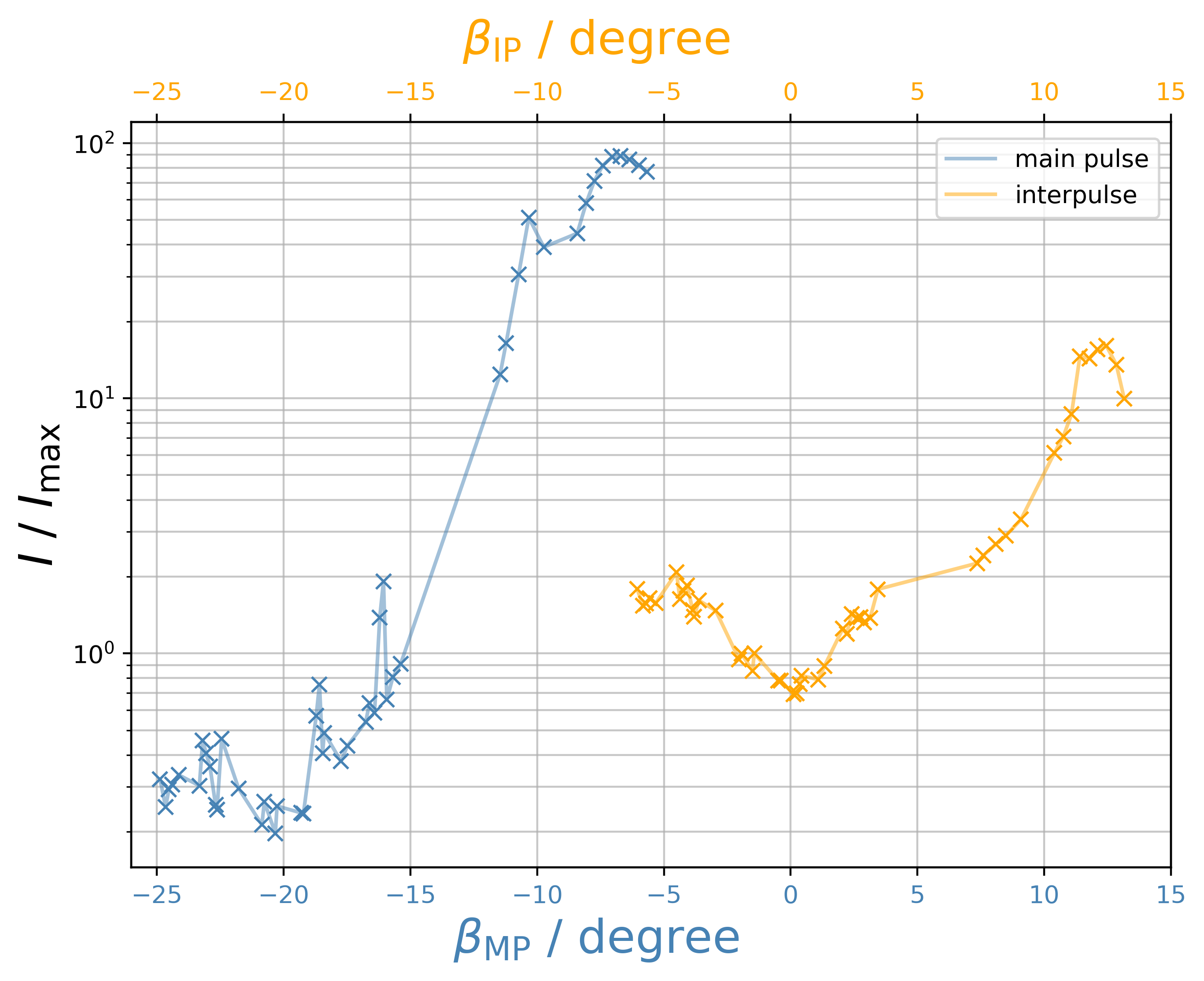}
	\caption{Latitudinal beam shape of the main pulse and interpulse of J1906$+$0476. Each point shows the maximum intensity of each beam in one of the 47 pulse profiles. The intensity is given in arbitrary units, as we are only interested in the relative intensities. $\beta_\mathrm{MP}$ denotes the value of $\beta$ measured with respect to the magnetic axis of the main pulse, $\beta_\mathrm{IP}$ denotes the value of $\beta$ measured with respect to the magnetic axis of the interpulse. Due to the precession of the pulsar, $\beta$ also can be read as a timeline from the most right point of every pulse to the left. 
	Uncertainties in the shown values can result from calibration errors or interstellar scintillation. We estimate them for each data point to be of the order of 10-20\% which we do not mark on this logarithmic scale.
	}
	\label{fig:pulseprofile}
\end{figure}

The angular separation of the LOS to the corresponding magnetic pole is designated as
$\beta$, which is usually constant for ordinary pulsars but here becomes a function of time due to precession. The various observed profiles consequently provide a cross section through the emission beams, similar to a tomography of the beam, across both magnetic poles. In order to describe the latitudinal structure, we measure the maximum intensity of each pulse (main pulse and interpulse, for each pole, separately) in every pulse profile and record these values for each of the 47 observations, freely available online\footnote{https://zenodo.org/record/3358819} \citep[see][]{Desvignes1013}, from 10\textsuperscript{th} July 2005 to the 21\textsuperscript{th} June 2018.
We do this by splitting the pulse profile recorded in 2048 phase bins into two parts,
for main and interpulse and treat them independently hereafter. 
 The geometry 
derived by \citet{Desvignes1013} from the \citet{kw09} precession model  allows us to assign each observation an impact angle $\beta(t)$.
Hereby we introduce $\beta_\mathrm{MP}$ and $\beta_\mathrm{IP}$ to indicate that the corresponting values of $\beta$ are measured with respect to the magnetic axis of the main pulse (MP) or interpulse (IP), respectively. 
As shown in Figure \ref{fig:pulseprofile}, the beams do not have 
a uniform shape but, in fact, we identify a number of sub-maxima in the main pulse.
The apparent rapid variation in some of the intensity values may be caused by calibration errors
or may be due to scintillation caused by the interstellar medium, although we cannot rule out
that they are real, reflecting a fine structure of the beam. We will later smooth over these
apparent short-term features, as they are irrelevant for the further study.

Overall, these unique data allow us to conclude already
that the general assumption of a uniform, box-shaped intensity distribution 
is not the best depiction of the beam shape.
We also note that the interpulse shows a clear intensity drop around $\beta_\mathrm{IP} = \SI{0}{\degree}$, i.e.
above the pole, which is expected from magnetospheric theories as pointed out by \cite{Desvignes1013}. It is also worth emphasizing
that the bright region of the main pulse is less wide than expected (see Introduction). Combined,
with this new perception about the intensity distribution of the pulsar beam, we conclude that we can improve on the modelling done by \cite{Pol_2019} by taking these observations into account.

\noindent

\subsection{Reconstructing the latitudinal beam shape}
\label{sec:reconstruction}

\begin{figure*}
	\centering
	\includegraphics[width=0.9\textwidth]{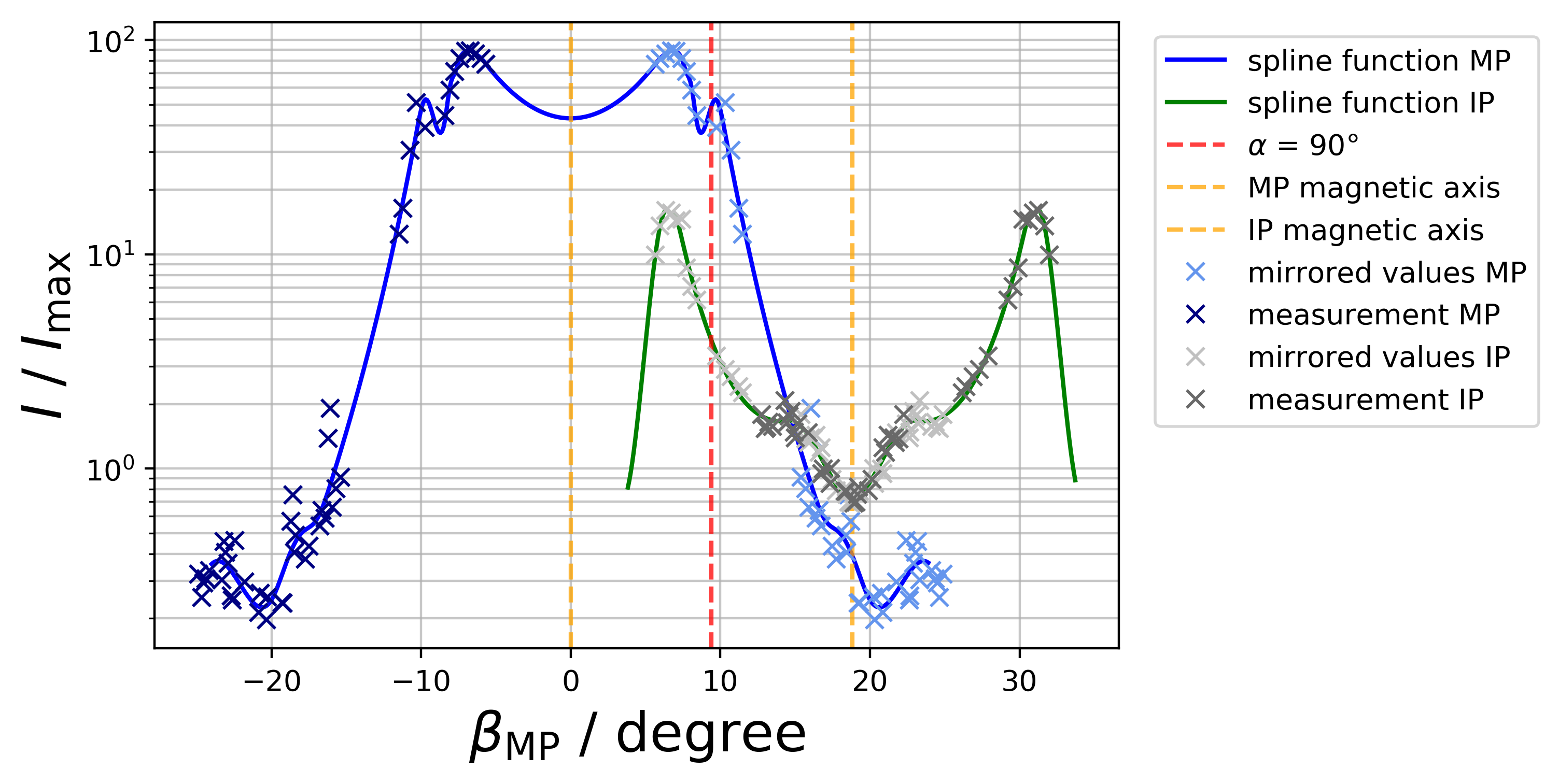}
	\caption{Similar to Figure~\ref{fig:pulseprofile}, showing the latitudinal beam intensity
	profile for both the main (blue) and interpulse (green), but performing a spline interpolation and mirroring the observed data at the location of the magnetic axis at each pole. For each pole, the darker shaded data points show the maximum intensities of the measured profiles. The lighter shaded data points indicate the mirrored values. The interpolating function used for each pulse is a cubic spline. $\beta_\mathrm{MP}$ is measured with respect to the main pulse magnetic axis, the intensity is given in arbitrary units, as we are only interested in relative intensities. The location of both magnetic axes
    is marked by yellow lines. The red line marks the pulsar's (rotational) equator, 
    i.e.\ $\alpha = \SI{90}{\degree}$.}
	\label{fig:composition}
\end{figure*}

Even though relativistic spin-precession provides access to the latitudinal beam structure,
the information is not complete. Due to gaps in the observations between 1998 and 2005, as well as between 2005 and 2009 \citep{Desvignes1013}, a reconstruction requires a limited set of assumptions
and extrapolation from the available data. 
This is done taking two conditions into account: Firstly, the pulsar has been detected with a strong signal in 1998 \citep{lsf+06}, where $\beta_\mathrm{MP}$ was greater than $\SI{0}{\degree}$  \citep{Desvignes1013}. Secondly, assuming that for an almost orthogonal rotator as we have here ($\alpha \sim {90}$ deg), we can
expect to first order an axial symmetry with respect to the magnetic axis. Based on these 
assumptions, we extrapolate the beam in form of mirroring intensity profiles of both main pulse and
interpulse along the axis with $\beta_\mathrm{MP/IP} = \SI{0}{\degree}$, the results which is shown in Fig.\ \ref{fig:composition}.  

Finally, in order to obtain a continuous intensity distribution as a function of $\beta$ for our later modelling, we perform a spline interpolation.  The lack of information about the main pulse's beam profile in the range of $\beta_\mathrm{MP} \in [-\SI{5}{\degree}, \SI{5}{\degree}]$ poses a major problem. We
apply the \texttt{LSQUnivariateSpline} function from the Python package \texttt{numpy.interpolate}, 
because it returns a spline function with the option of setting explicit internal knots, so the edges can be shaped manually. The different shapes of the main pulse and the interpulse impose a slightly different interpolation process for both beams.

Starting with the main pulse, the values for $\abs{\beta_\mathrm{MP}} > \SI{15}{\degree}$ show short-term
oscillations as discussed before (cf.\ Fig.\ \ref{fig:pulseprofile}). Hence, for this range
we set grid points manually. It turns out that grid points at $\pm\SI{16}{\degree}$, $\pm\SI{17}{\degree}$, $\pm\SI{18}{\degree}$, $\pm\SI{19}{\degree}$ and $\pm\SI{20}{\degree}$ yield a stable spline that represents the shape of the profile. 
Obviously, the distinct structures in the outer region of the profile can cause an instability of the interpolation, strongly depending on the position of the grid points. This restricts the number and position of grid points. To model the pulse around $\beta_\mathrm{MP} \in [-\SI{10}{\degree}, \SI{10}{\degree}]$, we identify the grid points in this interval with the actual data points to ensure that the spline follows the explicit shape of the measured beam as well as ensures 
a smooth interpolation of the profile in  $\beta_\mathrm{MP} \in
[-\SI{5}{\degree}, \SI{5}{\degree}]$. We are uncertain, as to whether the intensity peak at
$\beta_\mathrm{MP} \sim -10$ deg is a real feature of the pulse profiles. Judging from the slopes of the
intensity profile on either side, we are inclined to consider it as real and will therefore
include it in our modeling of the beam shape.


We applied the same scheme to the interpulse. The edge of the observed beam shape was modelled using manually chosen grid points (set to $\SI{-2}{\degree}$ to $\SI{3}{\degree}$ on the left side and $\SI{35}{\degree}$ to $\SI{40}{\degree}$ on the right side, both in steps of $\SI{1}{\degree}$.)
We show the results in Fig.\ \ref{fig:composition}, 
where the composition of both interpolated latitudinal beam shapes is depicted. 


\subsection{Implementation in the code framework}
\label{sec:implementation}

As shown in Fig.~\ref{fig:composition}, the measured latitudinal main pulse and interpulse beam shape differ significantly in their shape and relative intensities. This is in sharp contrast to the 
assumptions in 
previous studies, where one would compute the beaming correction factor based on
(measured or estimated) values for the angles $\alpha$ and $\rho$ according to
\begin{equation}
\label{eq:beamingfrac}
\frac{4\pi}{f_\mathrm{b}} = 2 \times \int_{0}^{2\pi}\mathrm{d}\phi \int_{\max(0, \alpha - \rho)}^{\min(\pi/2, \alpha + \rho)} \sin\theta \mathrm{d}\theta,
\end{equation}
which imposes the very simple beaming model of two equal beams with hard-edged cones (cf.\ \citealt{O_Shaughnessy_2010}). 
%

In the case of PSR J1906+0746 we can use the determined beam information directly.
The overlap of both profiles (main and interpulse) as well as the relative intensity to each other pulse needs to be handled with care, however, as it does not matter for detection statistics, if
the pulsar is detected by its main pulse, its interpulse, or both. For this reason, we construct an effective beam shape that reflects the illuminated sky, using the higher intensity of a given beam if the same spot is illuminated twice.

As we have modelled the intensity in a logarithmic scale to trace the beam edge more accurately, the intensity values obtained from the splines are exponentiated and afterwards normalised with respect to the maximum value of both poles, i.e.\ the maximum value of the main pulse spline. 

In our simulations, PSR J1906$+$0746-like pulsars will be assigned a luminosity that is drawn from an assumed distribution describing the whole population (see 
\citealt{Pol_2019}). We also draw 
a random value of $\zeta$ $\in [\SI{0}{\degree}, \SI{180}{\degree}]$, which 
is uniformly distributed in $\cos(\zeta)$, from which we derive $\beta_\mathrm{MP}$ as $\zeta = \alpha + \beta$. The observable radio intensity is that of the model luminosity scaled to a new value according to the value of $\beta_\mathrm{MP}$ and the inferred relative latitudinal intensity. 

By choosing a value of $\zeta$ along the whole polar angle, the fraction of the sky that is illuminated by the pulsar's beam is already intrinsically considered within our simulations. Hence,
when later analysing the results from the {\sc PsrPopPy}
simulations (see Section~\ref{sec:psrpoppy}), the beaming correction factor in Eqn.\ \ref{eq:mergerrate} is set to $f_\mathrm{b} = 1$.

\section{A generic latitudinal beam shape for other pulsars}

\label{sec:otherpulsars}

PSR J1906$+$0746 is not the only DNS system where relativistic spin precession has been observed.
Information on the beam structure is also available for the DNSs B1913+16 
\citep{kramer98,wt02} and
B1534+12 \citep{sta04}, and the relativistic NS-white dwarf system PSR J1141$-$6545 
\citep{mks+10,vbv+19}. In none of these cases, however, has the beam been
traversed completely yet, so that the important latitudinal extent cannot be determined as for PSR J1906+0746. (Only 
PSR J0737$-$3039B, the slow companion in the Double Pulsar \citep{lbk+04} has precessed out of our
line-of-sight \citep{pmk+10}, but here the beam is severely distorted by wind of pulsar A
\citep{mkl+04}.) What is common to all beam shapes observed so far
(including that of the non-recycled pulsar J1141$-$6545   
\citep{mks+10,vbv+19}), is that they do not seem to
have a uniform structure, as it is assumed in all previous studies to infer DNS merger rates
from the known population. We therefore improve on this assumption by utilizing the observed
beam structure of PSR J1906+0746 to derive a 
more realistic template for a beam shape of other DNSs to
provide a new estimate on the DNS merger rate.  Even though the individual beam shapes may
in reality still differ from this template, a study of the variations in the resulting merger
rates allows us to quantify the corresponding systematic uncertainty related to the beam shape
for the first time.


In order to derive a generic pulse profile,  we consider that we have observed different parts of the beam in both main pulse and interpulse of PSR J1906$+$0746. A suitable combination of these well measured portions in both beams describes the full latitudinal extent that is crucial for a realistic simulation of the pulsar detectability in a mock survey. We choose the edge of the main pulse as the edge of the generic beam shape. We use the centre 
of the interpulse to model the central part of the generic pulse shape.
In order to account for the gaps in the beam coverage as observation were not regularly spaced
(cf. Fig. \ref{fig:pulseprofile}), we make use of the spline-fitting discussed in Section~\ref{sec:reconstruction} and shown in Figure~\ref{fig:composition}. 

\begin{figure}
    \includegraphics[width=\columnwidth]{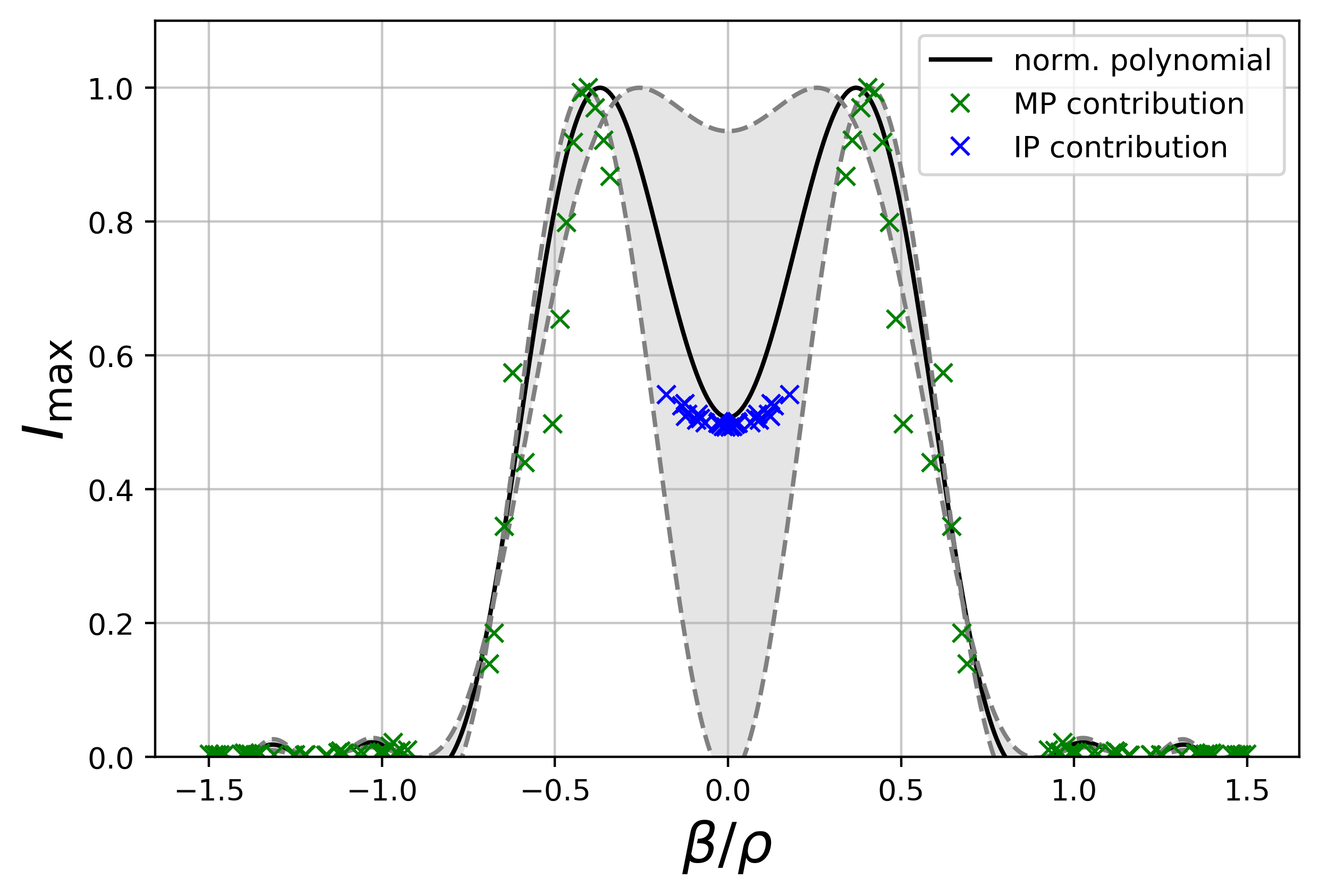}
    \caption{Exemplary composition of the generic latitudinal beam profile for one pulsar beam based on the observation of PSR J1906$+$0746 together with a polynomial fit function. Points marked in green are taken from the main pulse measurements, those marked blue belong to the interpulse measurements, shifted by an random offset of 0.45 in intensity. The fitted and normalized 
    polynomial representing the corresponding implemented profile is shown. The shaded region indicates the range of all possible other fitting functions for all different offsets. It is limited by the fit functions with the minimal and maximal offset of 0.0 and 0.9 for the interpulse points.}
    \label{fig:comp_generic_profile}
\end{figure}

The pulsar beam size, and hence its latitudinal extent, increases with decreasing period, $P$. As
discussed in the introduction, for dipolar field lines, one expects a scaling of 
$\rho = k \times P^{-0.5}$, which is confirmed statistically from observations. Given that
the factor $k$ has some uncertainty (see introduction and \cite{vbv+19}), we adopt the 
value of $k=5.4$ deg s$^{0.5}$ to be consistent with \cite{O_Shaughnessy_2010}. Computing the
corresponding $\rho$ for PSR J1906$+$0746, this implies a value of $\rho\sim14$ deg. 
This appears to be somewhat too small when inspecting Fig.~\ref{fig:pulseprofile}, but the intensity of the main pulse has dropped by a factor of $\sim 100$ at this value of $\beta$, so that
we consider this as a sufficiently good approximation, especially since all DNS studied
here have $P>20$ ms. We show the resulting latitudinal
beam shape after scaling the $\beta$-axis by a corresponding $\rho$ value in Figure~\ref{fig:comp_generic_profile}.

Before we apply the generic beam shape to simulated pulsars, we scale the profile 
in Figure~\ref{fig:comp_generic_profile} to the appropriate pulse period. We also 
vary it in a random fashion in two aspects. One is resulting from the uncertainty in
the intensity level of the clear minimum in the profile's centre upon the magnetic axis.
Here, for each simulated pulsar, we vary the relative intensity
at the centre of the beam by a random value in the range $[0.,0.9]$. The maximum value of 0.9 is
chosen in order to preserve at least a small intensity drop, as this is also expected 
theoretically \citep{ta13,glp17}.

The other aspect addresses the relative intensity of the beams above two magnetic poles.
From observations of interpulse pulsars (e.g.~\cite{jk19}) and from
an apparent lack of interpulse pulsars in the population of normal pulsars
(e.g.~\cite{wj08}), we deduce that the relative intensity between the beams may be remarkably different and may depend on some a priori unknown physical parameter.  We therefore assign
a random intensity ratio determined from two random draws from the log-normal
luminosity distribution that is used in the simulations. By definition, we call the 
resulting less luminous pulse the interpulse.

Finally, we assign every simulated pulsar a geometry, which consists of the angles 
$\alpha$ and $\beta$. For pulsars representing the realisations of DNSs with
known values of $\alpha$, we adopt those listed in Table \ref{tab:magnetic_angle}.
As $\alpha$ is yet unknown for the pulsars J0509+3801, J1757$-$1854, J1913+1102 and J1946+2052, for each simulated pulsar in the associated population we choose a random value distributed uniformly in $\cos({\alpha})$, consistent with $\zeta$ being uniformly distributed in $\cos({\zeta})$. In order to automatically account for the fraction of the sky illuminated by either main pulse or interpulse, we construct a combined profile as we had demonstrated for PSR J1906+0746 in Figure~\ref{fig:composition},
i.e.~we locate the main pulse at a position $\beta_\mathrm{MP} = \SI{0}{\degree}$ and centre the weaker interpulse accordingly at $\beta = 2 \times (\SI{90}{\degree} - \alpha)$.

\begin{table}
\caption{Values for the magnetic inclination angles applied in the implementation of the generic pulse profiles.}
    \centering
    \begin{tabular}{lrl}
    \hline
    \hline
    \noalign{\smallskip}
        Pulsar & $\alpha$ (deg) & Reference \\ \hline
            \noalign{\smallskip}
        J0737$-$3039A	& ${79}$ & \cite{ksv+21} \\
        B1534+12 & ${103}$ &  \cite{sta04}	\\
        J1756-2251 & ${74}$ & \cite{Ferdman_2014} \\
        J1906+0746 & ${81}$ & \cite{Desvignes1013} \\
        B1913+16 & ${153}$ & \cite{kramer98} \\
            \noalign{\smallskip} 
        \hline
    \end{tabular}

    \label{tab:magnetic_angle}
\end{table}

\section{Results}

In the following we apply our derived latitudinal beam shapes to simulations of the
Galactic DNS population and computations of the inferred DNS merger
rate as an enhancement of the work by \cite{Pol_2019,Pol_2020}. We do this in three steps. After first confirming
that we can reproduce the results by Pol et al., we then modify the simulations
by using only the new information on the beam shape  for PSR J1906$+$0476 to gauge the impact
of this change. In a third step, we then apply the newly derived 
generic latitudinal beam shape to all DNS systems, which allows us to compare
the different resulting merger rates and to assess the importance of these changes.

\subsection{Validating the simulation scheme}

Before applying the derived beam shape model, we use the information provided by Pol et al.\footnote{\url{https://github.com/NihanPol/2018-DNS-merger-rate/tree/master/python}} to 
compare our obtained results with those derived by  \citep{Pol_2019,Pol_2020}
(see Table~\ref{tab:resultspoletal}).
For most pulsars we could use the actual simulation and parameter files used by Pol et 
al.~for a detailed comparison. Only for PSR J0509+3801 added in \cite{Pol_2020} this was
not possible, and we derived our own input data. As a result, 
we derive a Galactic merger rate estimate of $\mathcal{R}^{\mathrm{Pol}}_{\mathrm{MW, recr}} = 37^{+24}_{-11}\ \si{\per\mega\yr}$ and an Advanced LIGO detection rate of $\mathcal{R}^{\mathrm{Pol}}_{\mathrm{LIGO, recr}} = 3.9^{+2.4}_{-1.2}\ \si{\per\yr}$,
which are a little smaller but in very good agreement with the values presented in \cite{Pol_2020}. The small deviation in the result are likely caused by statistical fluctuations in the simulation framework, the limited number of simulation runs in both cases, or some differences in the
input data (e.g.~the estimated lifetime of PSR J0509+3801). In the following, we 
use our derived value as the reference to compare to the changes when introducing 
improved beam shape modelling.

.

\subsection{Applying beam shape modelling for PSR J1906$+$0746}

Following the prescription described in Section \ref{sec:implementation}, the simulations
are repeated for all DNS sources but with modifications for
J1906+0746. We also apply the correction discussed in context of Eq.\ \ref{eq:ligorate} and a horizon distance of $\SI{130}{\mega\parsec}$. Following \cite{Pol_2020}, we also include
the recently discovered DNS J0509+3801 in our sample and include the 
corresponding Green Bank North Celestial Cap Survey (GBNCC) in our simulations \citep{Lynch_2018}.

\begin{figure*}
	\centering
	\includegraphics[width=0.8\textwidth]{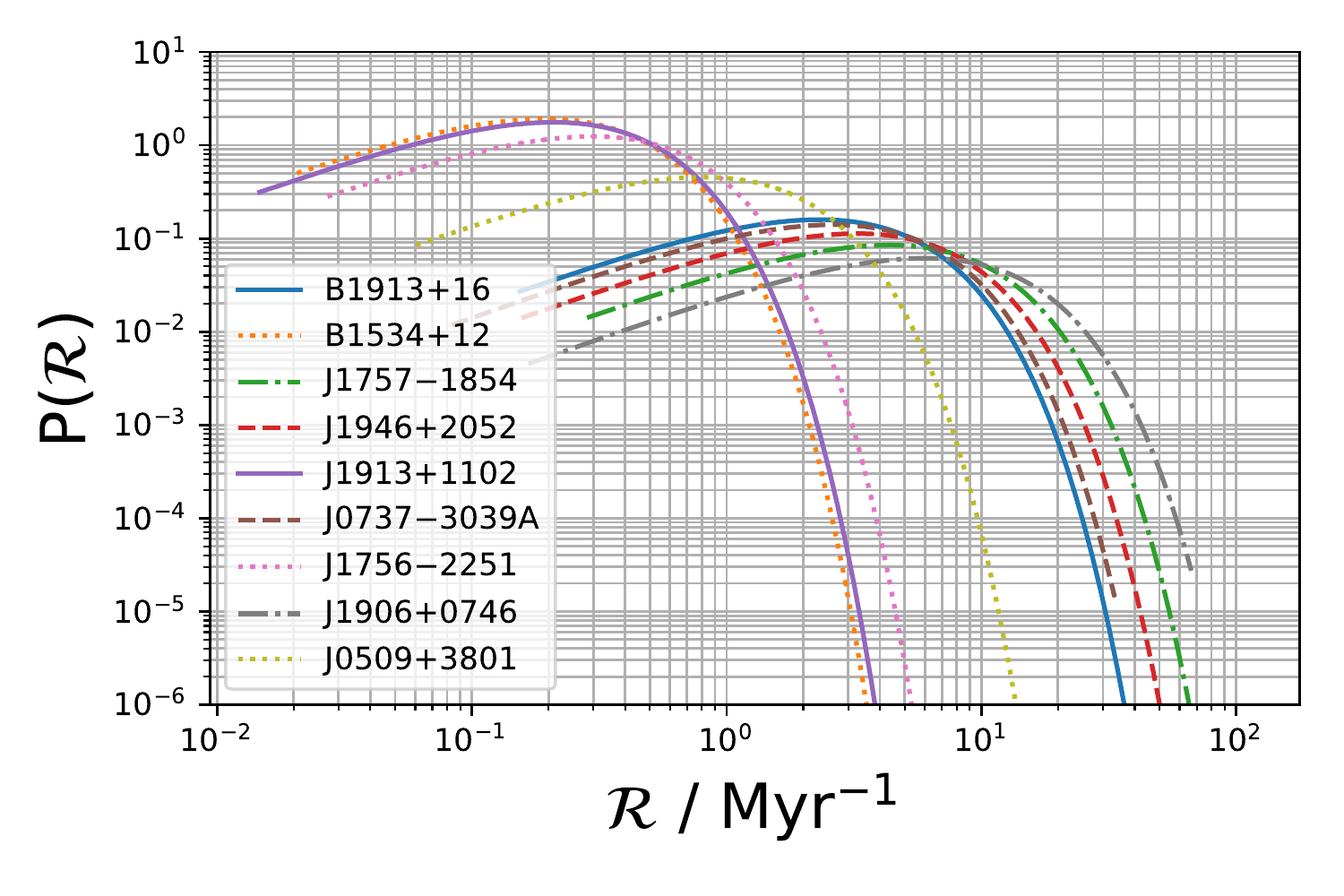}
	\caption{Galactic merger rate PDFs for the individual DNS systems. J1906+0746 is simulated using the extrapolated latitudinal beam profile (see Fig.\ \ref{fig:pulseprofile}). As result, the PDF for the J1906+0746-population extends to the largest rates in comparison.}
	\label{fig:newresindrate}
\end{figure*}
\begin{figure}
	\centering
	\includegraphics[width=\linewidth]{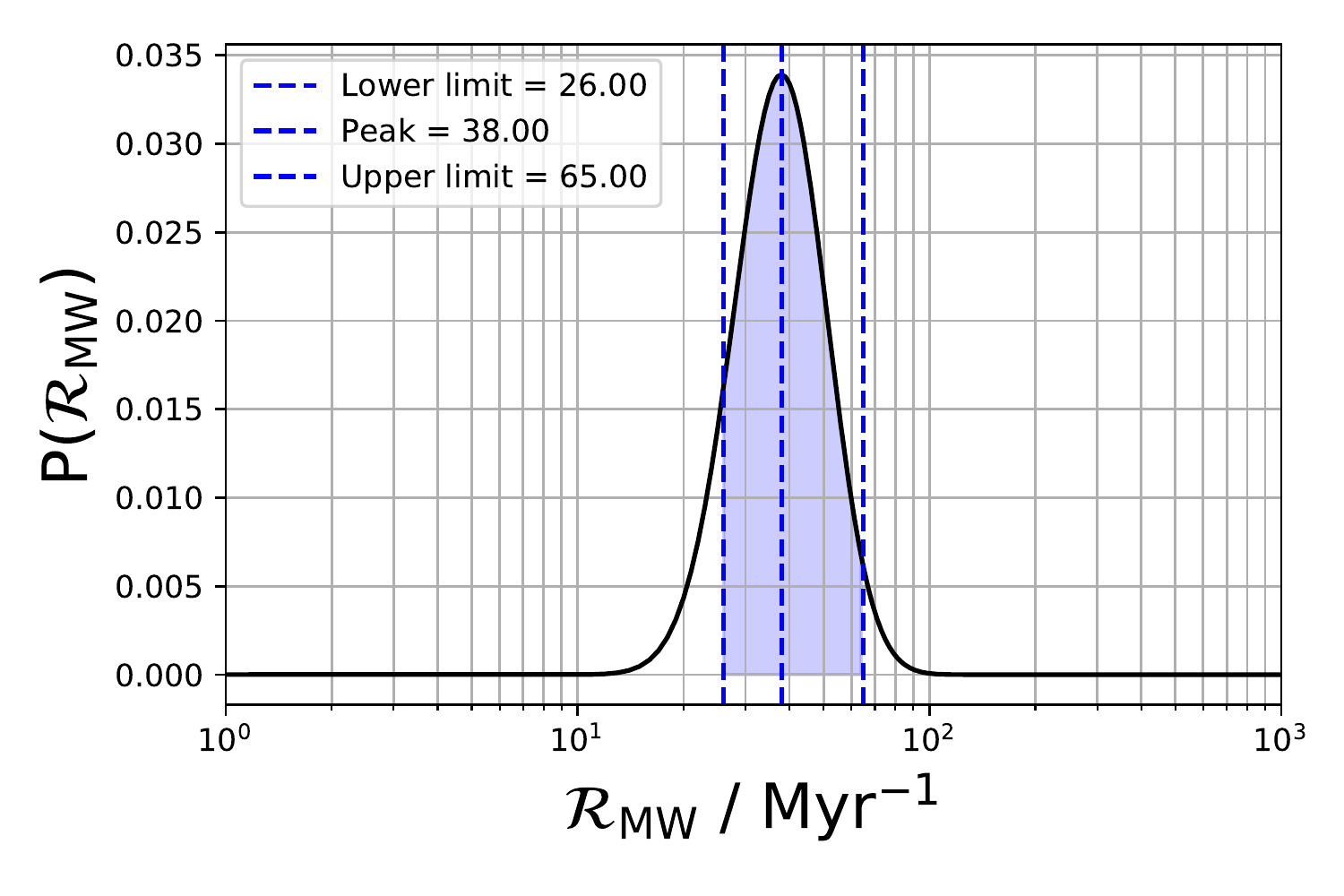}
	\includegraphics[width=\linewidth]{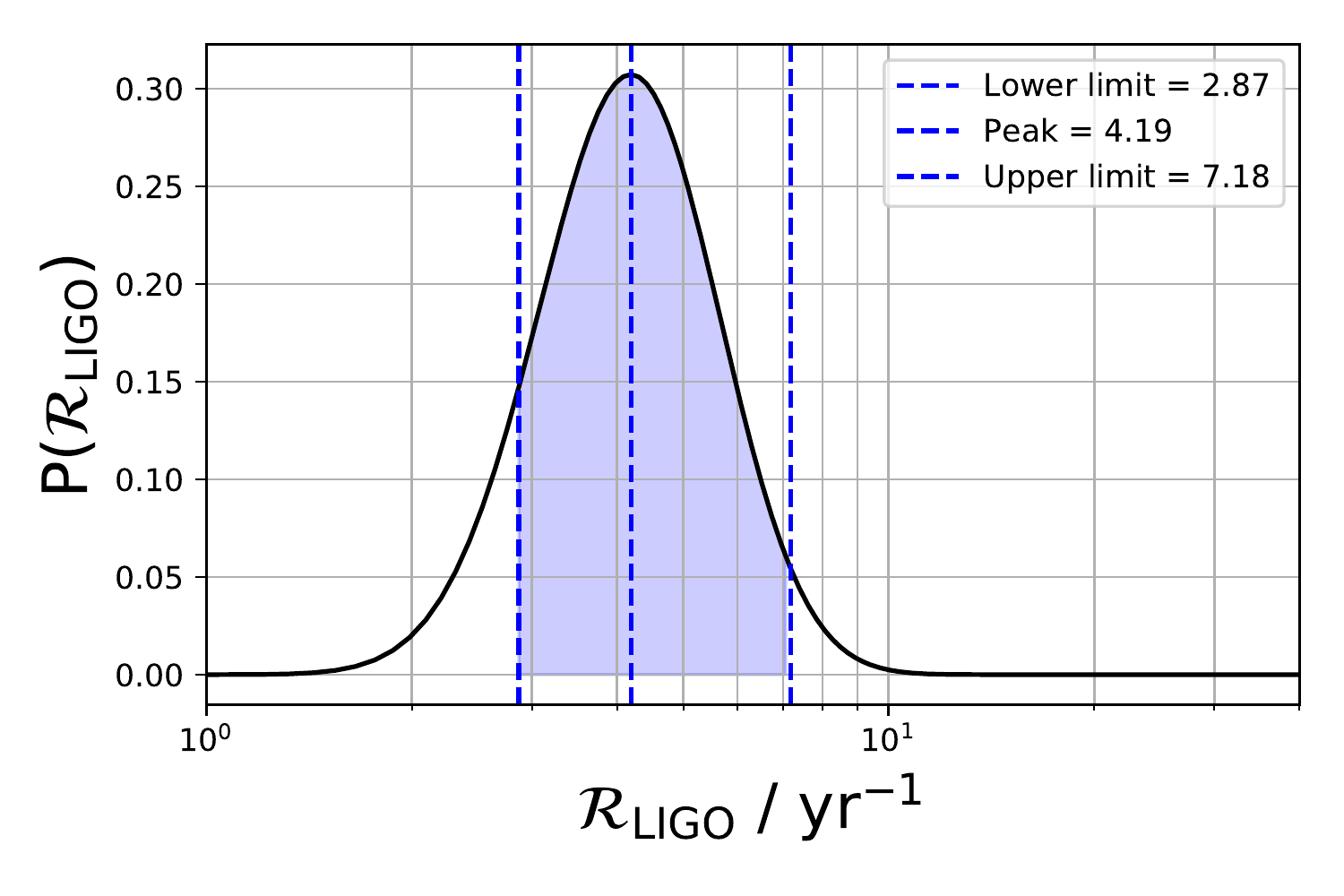}
	\caption{Estimates on the Milky Way DNS merger rate (top plot) and the DNS merger detection rate by LIGO (bottom plot) using the latitudinal beam profile (Fig.\ \ref{fig:pulseprofile}) for PSR J1906$+$0746. The errors are quoted at the 90\% confidence interval, with 45\% limits to the left and the right.}
	\label{fig:newresmergerrate}
\end{figure}\noindent

The results of our computations are shown in Figures \ref{fig:newresindrate} and \ref{fig:newresmergerrate}, where we can study the impact of introducing a more realistic
beam shape model. 
The impact of the new method is clearly seen in the increase of $N_{\mathrm{det}}$ from $56^{+236}_{-31}$ to $N_{\mathrm{det}}^{\mathrm{new}} = 361^{+1638}_{-266}$ (see also Figure~\ref{fig:rates1906}).
 But note that in our new scheme, the selection effect due to beaming is now fully addressed in our survey simulations, so that for PSR J1906$+$0746 we set 
$f_\mathrm{b} = 1$ in Eqn.~\ref{eq:ratewithfb}, therefore $N_{\mathrm{det}}^{\mathrm{new}} =  N_{\mathrm{tot}}^{\mathrm{new}} $. Thus, the peak value of the total population number increased 
very significantly, by 45\%.  The effect of the beam shape correction on the individual merger rate of J1906+0746 is shown in Fig.\ \ref{fig:newresindrate}: it shifts from $4.1^{+18.0}_{-2.4}\ \si{\per\mega\yr}$ to $\mathcal{R}^{\mathrm{new}} = 6.0^{+27.3}_{-4.5}\  \si{\per\mega\yr}$. This increase also affects the total Milky Way merger rate and the calculated LIGO merger rate. The new rate distributions are shown in both plots in Fig.\ \ref{fig:newresmergerrate}. The updated estimate on the Milky Way merger rate is  $\mathcal{R}^{\mathrm{new}}_{\mathrm{MW}} = 38^{+27}_{-12}\ \si{\per\mega\yr}$. The new DNS merger rate estimate for LIGO resulting from the changes applied to PSR~J1906$+$0746 is \begin{equation}
    \mathcal{R}_{\mathrm{LIGO}}^{\mathrm{new,1906}} = 4.19^{+2.98}_{-1.32}\ \si{\per\yr}
\end{equation}
which imposes indeed a significant increase. The errors of both values are quoted at the 90\% confidence level.

\begin{table*}
\begin{threeparttable}
	\caption{Parameters and results from the simulation using the values of $f_\mathrm{b}$ adopted by Pol et al.~(2019).}
	\label{tab:resultspoletal}
	\begin{tabularx}{\textwidth}{lXXXXXXX}
	\hline
	\hline
	\noalign{\medskip}
		Pulsar & $f_\mathrm{b}$ & $\delta$ & $\tau_{\mathrm{age}}$ & $t_\mathrm{merger}$ & $\gamma$ & $N_{\mathrm{tot}}$ & $\mathcal{R}$ \\
		&&& (Myr) & (Gyr)&&& (Myr$^{-1}$)\\ 
		\noalign{\medskip}
		\hline 
		\noalign{\medskip}
		J0509+3801 & 4.59 & 0.18& & & 0.007437 & $614^{+2823}_{-422}$ & $0.8^{+3.7}_{-0.6}$\\
		&&&&&&&\\ [-1em]
		J0737$-$3039A & 2.0 & 0.27& 159 & 0.085 & 0.003183 & $627^{+2868}_{-463}$ & $2.6^{+11.9}_{-1.9}$\\
		&&&&&&&\\ [-1em]
		J0737$-$3039B\footnotemark& & & & 0.085 & & &\\
		&&&&&&&\\ [-1em]
		B1534+12 & 6.0 & 0.04 & 208 & 2.70 & 0.010687 & $567^{+2581}_{-386}$ & $0.2^{+0.9}_{-0.1}$\\
		&&&&&&&\\ [-1em]
		J1756-2251 & 4.59 & 0.03 & 396 & 1.69 & 0.009133 & $504^{+2292}_{-348}$ & $0.3^{+1.4}_{-0.2}$ \\
		&&&&&&&\\ [-1em]
		J1757$-$1854 & 4.59 & 0.06 & 159 & 0.076 & 0.006522 & $706^{+3208}_{-495}$ & $4.4^{+19.8}_{-3.1}$\\
		&&&&&&&\\ [-1em]
		J1906+0746 (old) & 4.59 & 0.01 & 0.11 & 0.30 & 0.019134 & $248^{+1082}_{-147}$ & $4.1^{+18.0}_{-2.4}$\\
		&&&&&&&\\ [-1em]
		J1906+0746 (new) & 1 & 0.01 & 0.11 & 0.30 & 0.002781 & $361^{+1638}_{-266}$ & $6.0^{+27.3}_{-4.5}$\\
		&&&&&&&\\ [-1em]
		J1913+16 & 5.7 & 0.169 & 77 & 0.50 & 0.006681 & $651^{+2998}_{-458}$ & $2.3^{+10.6}_{-1.7}$\\
		&&&&&&&\\ [-1em]
		J1913+1102 & 4.59 & 0.06 & 2625 & 0.30 & 0.007019 & $816^{+3777}_{-586}$ & $0.2^{+1.0}_{-0.1}$\\
		&&&&&&&\\ [-1em]
		J1946+2052 & 4.59 & 0.06 & 247 & 0.046 & 0.004800 & $963^{+4363}_{-697}$ &$3.3^{+14.9}_{-2.4}$ \\
		\noalign{\medskip}
		\hline
	\end{tabularx}

	\begin{tablenotes}
	\item \textit{Note 1}$f_\mathrm{b}$ is the beaming correction factor, $\delta$ the pulse duty cycle and $\tau_{\mathrm{age}}$ is the effective age of the pulsar. The numbers given for $N_{\mathrm{det}}$, $N_{\mathrm{tot}}$ and $\mathcal{R}$ denote the peak values resulting from the probability distributions \ref{eq:ntotrate} and \ref{eq:mergerrate}. $N_{\mathrm{det}}$ is the number of pulsars beaming towards the earth, $N_{\mathrm{tot}}$ is the total population number and $\mathcal{R}$ is the individual merger rate. $N_{\mathrm{tot}}$ is gained by scaling $N_{\mathrm{det}}$ with $f_\mathrm{b}$.
	\item \textit{Note 2} To remain consistency with the work by \protect\cite{Pol_2019}, all errors are quoted on the 95 \% confidence intervals.
	\item \textit{Note 3} PSR J1906+0746 (old) denotes the results of the simulation based on the concept by Pol et al., J1906+0746 (new) denotes the results of the simulation using the spline interpolated latitudinal intensity profile.  
	\end{tablenotes}
\end{threeparttable}
\end{table*}

\footnotetext{The J0737-3039 system was discovered with pulsar A, also pulsar B will cross the death line significantly earlier than pulsar A. In addition to that, pulsar B introduces large uncertainties into the total merger rate \cite{Pol_2019}. Therefore it will not be taken into account for the analysis.}


\subsection{Effective beaming correction factor}
\label{ssec:fbeff}

Using the PDFs of the merger rates generated from the generic latitudinal profile simulation, it is possible to determine an effective beaming correction factor $\Tilde{f}_\mathrm{b,eff}$. We define it as the value for $f_\mathrm{b}$, at which the merger rate PDF obtained from the simulation following the procedure from \cite{Pol_2019} coincides with the merger rate PDF from the simulation using generic pulse profile, i.e. $\Tilde{f}_\mathrm{b,eff} = \gamma_\mathrm{old}/\gamma_\mathrm{new}$.
$\Tilde{f}_\mathrm{b,eff}$ is iteratively determined by imposing a maximal difference between the maxima of the two PDFs. The individual results are shown in the last column of Table~\ref{tab:resultsgeneric}. For PSR~J1906$+$0746 we apply an alternative way to compute $\Tilde{f}_\mathrm{b,eff}$ by using using our spline interpolation of the beam shape directly. 
This results in $\Tilde{f}_\mathrm{b,eff} = 6.98$. This is significantly larger than the value adopted by \citet{O_Shaughnessy_2010} (${f}_\mathrm{b,eff} = 3.37$) due to the lack of additional information, or the uniform value adopted
by \citet{Pol_2019} (${f}_\mathrm{b,eff} = 4.60$). The impact of adopting an
appropriate beam shape is clearly demonstrated.

\subsection{General application of the generic latitudinal beam shape}

\begin{figure*}
	\centering
	\includegraphics[width=0.8\textwidth]{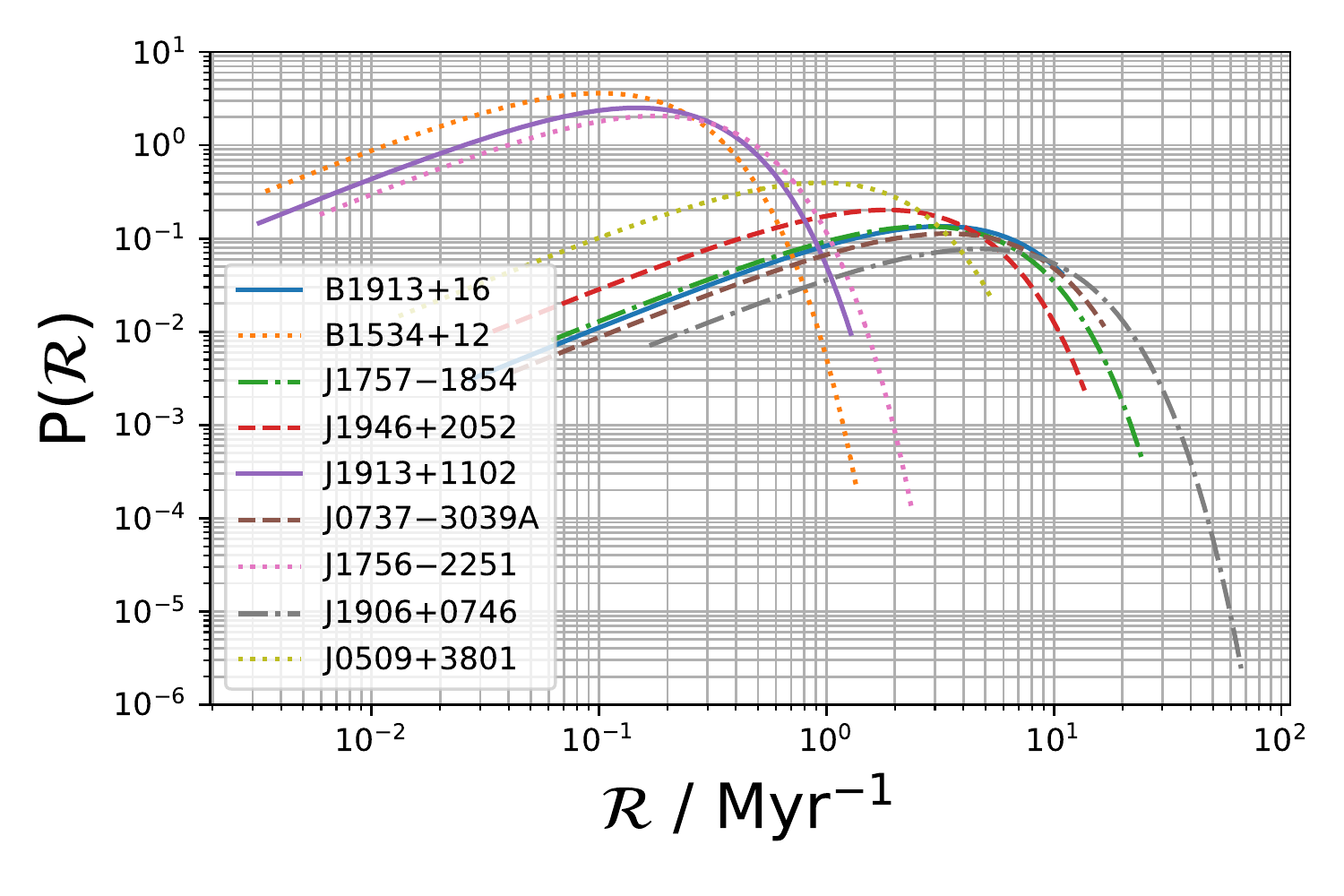}
	\caption{Galactic merger rate PDFs for the individual DNS systems based on the simulation using the generic latitudinal beam profile. We still encounter the accumulation of two distinct groups in terms of the PDF's relative position, yet overall the curves shifted to the left, i.e.\ to smaller peak positions.}
	\label{fig:ratesgeneric}
\end{figure*}

In the last step we adopt the generic beam shape mode for all pulsars listed in Tab.\ \ref{tab:resultsgeneric} following the procedure described in Sec.\ \ref{sec:otherpulsars}. Again, for the extrapolation to the LIGO DNS detection rate, we use a horizon distance of $\SI{130}{\mega\parsec}$. The numerical results are listed in Tab.\ \ref{tab:resultsgeneric}. Comparing the values of $N_\mathrm{det}$ from the previous simulation to the simulation with the generic profile, we firstly see a significant increase in the number of all DNS systems. However,
when computing the merger rates (hence using $f_\mathrm{b}=1$ in Eqn.~(\ref{eq:ratewithfb}) 
for all pulsars), 
most contributions to the overall rates decreased. The individual merger rates following from the simulation are shown in Fig.\ \ref{fig:ratesgeneric}. Inspecting the location of the peaks in the DNS merger rates, compared to the earlier reproduction of the Pol et al.\ results (Fig.\ \ref{fig:newresindrate}), we can see that most of the curves moved to slightly lower
values. Not surprisingly, we encounter the same grouping of curves: The three pulsars B1534$+$12, J1913$+$1102 and J1756$-$2251 still contribute least to the overall DNS merger rate, but yet the other PDF curves appear closer to each other. The PDF from the PSR J0509$+$3801 population merger rate moved into the right group of curves and also the position of the individual PDF curves changed within that group. Importantly, PSR J1906$+$0746 represents the DNS population with the biggest contribution to the galactic DNS merger rate, which is indeed to be expected given its young age.
The contributions of PSRs J0737$-$3039A and J1913$+$1102 increased noticeably in relation to the other curves. The resulting estimate on the Milky Way DNS merger rate is $\mathcal{R}^{\mathrm{gen}}_{\mathrm{MW}} = 32^{+19}_{-9}\ \si{\per\mega\yr}$. The estimate for the LIGO detection rate becomes
\begin{equation}
    \mathcal{R}^{\mathrm{gen}}_{\mathrm{LIGO}} = 3.5^{+2.1}_{-1.0}\ \si{\per\mega\yr}
\end{equation}
which represents a decrease in comparison to the results of \cite{Pol_2020}, but is still consistent with their values. Again the errors are quoted at the 90\% confidence level.

\begin{table}
\begin{threeparttable}
    \centering
	\caption{Parameters and results from the simulation using the generic latitudinal pulse profile. The errors are quoted at the 90\% confidence interval.}
	\label{tab:resultsgeneric}
	\begin{tabular}{llrrr}
	\hline
	\hline
	\noalign{\medskip}
		Pulsar &  \multicolumn{1}{c}{$\gamma$} &
		$N_{\mathrm{det,tot}}$
		& \multicolumn{1}{c}{$\mathcal{R}$} & $f_\mathrm{b, eff}$ \\
		&&& (Myr$^{-1}$) &\\ 
			\noalign{\medskip}
		\hline
			\noalign{\medskip}
		J0509+3801 & 0.00138568 & $720^{+2700}_{-543}$ & $0.9^{+3.6}_{-0.7}$ & 5.24\\
		&&&&\\ [-1em]
		J0737$-$3039A & 0.0012233 & $817^{+2752}_{-619}$ & $3.4^{+11.5}_{-2.6}$ & 2.47 \\
		&&&&\\ [-1em]
		B1534+12 & 0.00334478 & $298^{+1366}_{-220}$ & $0.1^{+0.5}_{-0.1}$ & 3.20 \\
		&&&&\\ [-1em]
		J1756-2251 & 0.00332935 & $302^{+1370}_{-224}$ & $0.2^{+0.8}_{-0.1}$  & 2.76\\
		&&&&\\ [-1em]
		J1757$-$1854 & 0.0022562 & $441^{+2001}_{-328}$ & $2.7^{+12.4}_{-2.0}$ & 2.87 \\
		&&&&\\ [-1em]
		J1906+0746 & 0.00350737 & $286^{+1302}_{-208}$ & $4.8^{+21.7}_{-3.5}$ & 5.50\\
		&&&&\\ [-1em]
		J1913+16 & 0.00084191 & $1188^{+2604}_{-919}$ & $3.2^{+7.0}_{-2.5}$ & 6.71\\
		&&&&\\ [-1em]
		J1913+1102 & 0.00217771 & $457^{+2065}_{-339}$ & $0.1^{+0.7}_{-0.1}$ & 3.21\\
		&&&&\\ [-1em]
		J1946+2052 & 0.00187952 & $533^{+2325}_{-399}$ & $1.8^{+7.9}_{-1.4}$ & 2.55 \\
			\noalign{\medskip}
			\hline
	\end{tabular}
    \begin{tablenotes}
    \item \textit{Note 1} $f_\mathrm{b, eff}$ is the effective beaming correction factor calculated as described in \ref{ssec:fbeff}. The numbers given for $N_{\mathrm{det, tot}}$ and $\mathcal{R}$ denote the peak values resulting from the probability distributions \ref{eq:ntotrate} and \ref{eq:mergerrate}. $N_{\mathrm{det}}$ is the number of pulsars beaming towards the earth, $N_{\mathrm{tot}}$ is the total population number and $\mathcal{R}$ is the individual merger rate. $N_{\mathrm{tot}}$ is gained by scaling $N_{\mathrm{det}}$ with $f_\mathrm{b}$. Since $f_\mathrm{b}= 1 $ for all pulsars, the distributions do not change from $N_{\mathrm{det}}$ to $N_{\mathrm{tot}}$ and thus the values are given in the same column.
    \item \textit{Note 2} To remain consistency with the work by \protect\cite{Pol_2019}, the errors on the population numbers are quoted on the 95 \% confidence intervals, whereas the errors on the merger rates are given with the 90 \% confidence interval.
    \end{tablenotes}
\end{threeparttable}
\end{table}

\section{Discussion}

There are a number of caveats and systematic uncertainties in the applied method and
framework. 
The caveats concerning the applied log-normal luminosity distribution used in the PsrPopPy2 simulation, the beaming correction factors, the effective lifetime of J1906+0746 and the extrapolation of the merger rate to the observable volume of LIGO outlined in Sec.\ 4.2 of the paper by \cite{Pol_2019} are equally valid in this work. However, since they have 
been discussed in detail in the cited work, we mention them only for completeness at this point
and refer to \cite{Pol_2019} for more details.

One may also consider whether all systems studied here represent ``true'' DNS systems, or whether an unseen companion may be indeed a heavy white dwarf. In principle, such possibility cannot be excluded. In the Double Pulsar, we have seen
both components of the system. In the case of PSR J1906$+$0746, we observe the non-recycled pulsar. It is possible to find a young pulsar around a white dwarf companion, as in the case of PSR J1141$-$6545 \citep{abw+11}, but such evolution requires finely tuned initial system parameters (e.g.~\cite{ts00}). In all other cases, the observed pulsar is recycled, and all systems have a significant eccentricity which indicate a second supernova explosion with an asymmetric kick and/or significant mass loss, strongly suggesting that these systems are DNSs.

It has recently been suggested that a population of massive radio-quiet neutron stars in compact binary systems could exist \citep{vsr+21}. In such a case, our derived LIGO detection rates will represent a lower-limit. If there is indeed a large difference to the observed detection rates, compared to our estimates here, it would lend credibility to such conclusion. However, one should note that so far there is no evidence for a correlation between pulsar luminosity and pulsar mass \citep{lk05}.

In the following we discuss further caveats
associated with the specifics of this work.

\subsection{Caveats on the latitudinal beam profile}

\subsubsection{Extrapolation from the PSR J1906$+$0746 measurements}

The main uncertainty in deriving the latitudinal beam profile
arises from the extrapolation of the main pulse. A fraction less than half of both latitudinal profiles is covered by the available observations and only the detection in 1998 gives information back in time to tie the extrapolation to. As no further measurements are available, the method of mirroring the beam shape is the only way to treat the pulse intensity profile. Therefore the actual shape of the latitudinal profile could vary significantly from the one assumed in this work. During our simulations, we have seen that a variation of the opening angle strongly affects the merger rate estimate, so that a different pulse shape could lead to overall different results in the individual contributions of the populations to the combined galactic merger rate. In addition, the lack of data points for both main pulse and interpulse from 2009 to 2012 forces us to interpolate 
the resulting wide gap nearly linearly as first-order approximation. The real pulse shape may be
different in this gap. As discussed previously, also the shape of the intensity distribution in the center of each pulse poses an uncertainty to our merger rate estimate for the population of PSR J1906$+$0746. 
A more accurate profile could lead to a larger illuminated fraction of the sky and thus to a small decrease of the DNS merger rate estimate.

\subsubsection{Validity of the generic beam shape }

The derivation of the generic profile from the profile of PSR~J1906$+$0746 
also bears some uncertainties, which can affect the resulting populations and thus the DNS merger rate estimate significantly. 

We chose the contribution of both magentic poles of PSR J1906$+$0746 to the generic profile in such a way, that the resulting polynomial depicts the observed shape the most suitable way. But surely we could approach this in many different ways, especially the choice of the individual contributions. For example, one could add up both profiles and fit a suitable function accordingly. But taking a closer look at Fig. \ref{fig:pulseprofile}, this would lead to a smeared shape, as the edges of both profiles do not coincide. Compensating for that by re-scaling parts of either the main pulse or the interpulse along the x-axis, one would have had to make alternative
assumptions about the rescaling magnitude and also which part to rescale.

Yet another way of gaining a generic profile would have been working only with the interpulse data, as via mirroring the values, ending up with an almost complete profile. This option also hides two peculiarities. At first, the value of $\beta_\mathrm{IP}$ corresponding to the maximum of the interpulse is larger than the value of $\beta_\mathrm{MP}$ corresponding to the maximum of the main pulse. Therefore taking the two main peaks from the latitudinal interpulse profile (cf.\ Fig.~\ref{fig:pulseprofile}) as maxima for the generic pulse profile would lead to a slightly larger latitudinal extent of the same. This in turn leads to smaller population sizes, probably resulting in an underestimation of the contribution to the DNS merger rate. Especially pulsar populations with large pulse widths are prone to that. Secondly, in order to take the center variations of the profile into account, one would have to decide arbitrarily, which data points define the "center" of the profile.

Taking all this into account, we decided following the approach pointed out in Sec.~\ref{sec:otherpulsars}. The most uncertain aspect of that strategy is the negligence of the plateaus near the middle of the interpulse, motivated by the circumstance that they caused numerical instabilities in the fitting process. The intensity level of the central values of the intensity profile above the pole, as well as the relative intensity of interpulse and main pulse also bear some uncertainty. However, given that
PSR~J1906$+$0746 is the only pulsar for which such a measurement has been possible so far, we believe our approach serves as a useful first-order approximation, improving upon the assumption of a uniformly filled beam which clearly contradicts the observations.


\subsection{Comparison of methods for PSR J1906$+$0746}

\begin{figure}
    \centering
    \includegraphics[width=\linewidth]{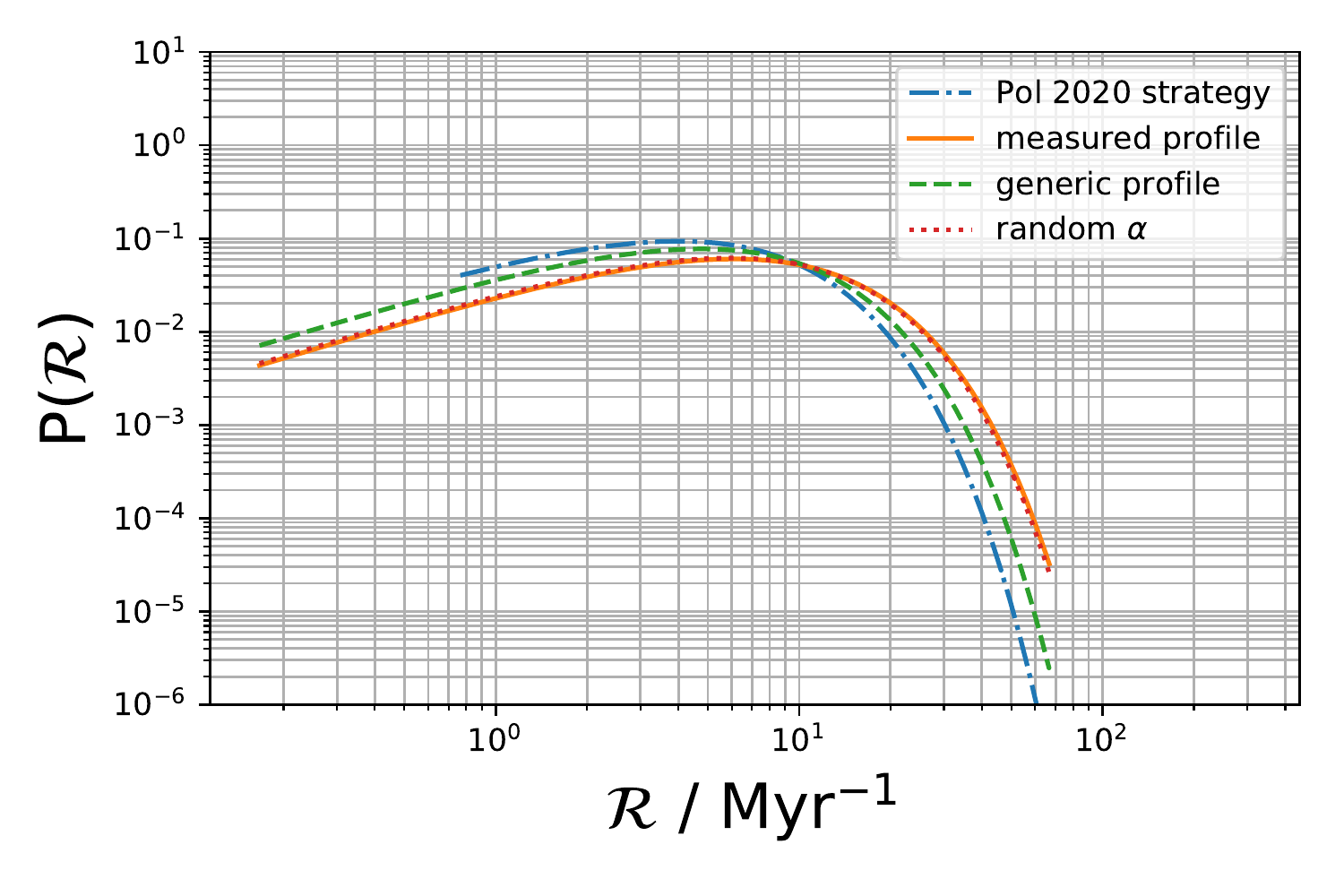}
    \caption{Comparison of the merger rates derived for PSR J1906$+$0746 using the variation of methods discussed in this work. The curves correspond in their order as they appear in the labelling to the different simulation strategies (i) to (iv) enumerated in Section \ref{enum:strategies_1906}.}
    \label{fig:rates1906}
\end{figure}

Our main focus in this work are the implications of the recent 
observational results for PSR~J1906$+$0746. In addition, for PSR J1906$+$0746, we have the unique opportunity to study the impact corresponding to the additional degrees of freedom introduced during the creation of the generic profile. This leads to a total of four different simulations on PSR J1906$+$0746 that we conducted, ie. those
\begin{enumerate}
\label{enum:strategies_1906}
    \item following the strategy of \cite{Pol_2019}
    \item applying the profile extracted from the observations
    \item applying the generic profile\footnote{\label{note:gen_prof}variation of the center intensity values and the intensity ration between the MP and IP allowed} using the known (fixed) $\alpha$
    \item applying the generic profile\footnote{See footnote \ref{note:gen_prof}.} and assuming $\alpha$ to be unknown, i.e.\ allowing it to vary.
\end{enumerate}
The results of all four simulations are shown in Fig.~\ref{fig:rates1906}, the values of $\gamma$ and $N_\mathrm{obs}$ are collected in Tab.~\ref{tab:1906_result}.
We see that the rates resulting from simulations (ii)-(iv) are shifted to the right with respect to the merger rate obtained from the initial simulation by \cite{Pol_2020}, meaning that in all three cases the peak merger rate increased. Nevertheless, we also find that the usage of the generic profile leads to a noticeably smaller increase than the usage of the observed profile.

Obviously, allowing the actually observed profile to vary in a random fashion when applying the generic profile has a noticeable impact. Due to the variation in the intensity ratio between main and interpulse, as well as varying the central intensity levels, we create, on average, a pulsar beam that is brighter than the original one. As a result, a corresponding population of pulsars is detected more often in our virtual pulsar surveys than otherwise, leading to a decrease in the merger rate estimate. Overall, however, we can conclude that it is crucial for the simulation to take the latitudinal beam shape into consideration.
\begin{table}
    \caption{Collection of all numerical results concerning the simulations on J1906+0746. These values are used to estimate the errors on the simulation of the other pulsars with unknown physical beam shape}
    \centering
    \begin{tabular}{lcccc}
    \hline
    \hline
    \noalign{\medskip}
         & $N_\mathrm{tot}$ & $\mathcal{R} (\si{\per\mega\yr}$)\\
     \noalign{\medskip}      
         \hline
           \noalign{\medskip}
        Pol et al. (2019, 2020) & $248^{+1082}_{-147}$ & $4.13^{+18.03}_{-2.44}$ \\
        &&&&\\ [-1em]
        measured profile & $365^{+1669}_{-272}$ & $6.09^{+27.82}_{-4.53}$ \\
        &&&&\\ [-1em]
        generic profile, fixed $\alpha$ & $286^{+1306}_{-208}$ & $4.76^{+21.77}_{-3.46}$ \\
        &&&&\\ [-1em]
        generic profile, random $\alpha$ & $357^{+1630}_{-264}$ & $5.96^{+27.16}_{-4.39}$ \\
          \noalign{\medskip}
          \hline
    \end{tabular}
    \label{tab:1906_result}
\end{table}


\subsection{Comparison with other DNS merger rate estimates}
\label{subsec:comparison}

The estimate for the rate of DNS merger detections by LIGO derived in this work can be compared to the ones obtained through different theoretical or phenomenological models. In comparison to the method used in this work, i.e.\ extrapolating the merger rate on the basis of the detected DNS systems, it is also possible to generate a population of Galactic DNS systems ab initio. Here the different (stellar and binary evolution) formation processes towards a DNS system are 
considered. There is a rich literature on this topic. Here, as an recent examples, 
we compare our results to the rate predictions by \cite{chruslinska} and by \cite{Kruckow_2018} shown in Fig.~\ref{fig:DNSmergercompilation}. We also consider the DNS merger rate 
calculated by the LIGO collaboration based on the observational input from
GW170817 and GW190425.

\subsubsection{LIGO DNS merger rate}

Due to the detection of two DNS mergers, GW170817 and GW190425, the LIGO collaboration released a new estimate on the DNS merger rate based on both these events \citep{Abbott_2020}. 
After the unit conversion as in \cite{Pol_2020}, we find
\[\mathcal{R}_{\mathrm{LIGO}} = 4.6^{+7.1}_{-3.4} \times \left(\frac{D_{r}}{\SI{100}{\mega\parsec}}\right)^3 \si{\per\yr}\]
Applying a luminosity distance of $D_r = \SI{130}{\mega\parsec}$, this gives a rate of $\mathcal{R}_{\mathrm{LIGO}} = 10.11^{+15.60}_{-7.47}\ \si{\per\yr}$, where the errors are quoted at the 90\% confidence interval. This rate is also plotted in Fig.~\ref{fig:DNSmergercompilation}.

\subsubsection{Ab initio simulations}

\cite{chruslinska} predicted a DNS merger rate density of $\SI{48.4}{\per\cubic\giga\parsec\per\yr}$ that translates to a rate of $\SI{0.1063}{\per\yr}$ using a LIGO range distance of $\SI{130}{\mega\parsec}$. This is clearly lower than the range of merger rates presented in this work. They also offer a variety of different models, where the most optimistic one derives a rate density of $600^{+600}_{-300}\ \si{\per\cubic\giga\parsec\per\yr}$, which corresponds to a merger rate range of $1.32^{+1.32}_{0.66}\ \si{\per\yr}$ using the previously introduced range distance.

\begin{figure*}
    \centering
    \includegraphics[width=\linewidth]{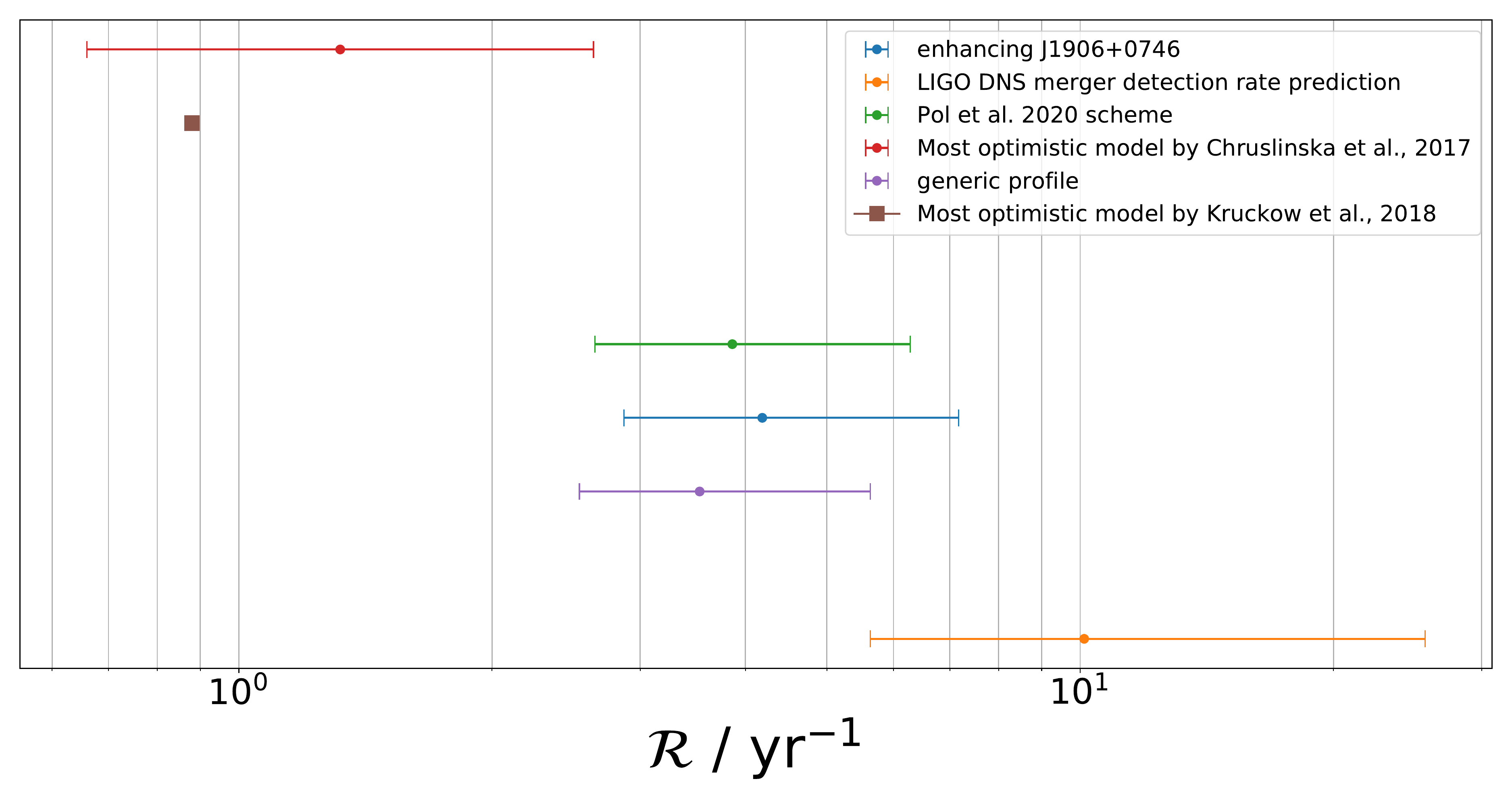}
    \caption{Comparison of the DNS merger rate estimates calculated in this work to
    previous work using different methods. These are the ab-initio estimates by \protect\cite{chruslinska} and \protect\cite{Kruckow_2018}, the LIGO estimate by \protect\cite{Abbott_2020} and the estimate from the simulation following the \protect\cite{Pol_2020} scheme (which is slightly smaller here than the value dervied by \protect\cite{Pol_2020} due to statistical fluctuations). Both estimates derived in this work are consistent with each other, as well as with the rate from \protect\cite{Pol_2020}.}
    \label{fig:DNSmergercompilation}
\end{figure*}

In comparison, \cite{Kruckow_2018} synthesised binary populations at different metallicities (Z$_\mathrm{MW}$ = 0.0088; Z$_\mathrm{IZw18}$ = 0.0002), taking into account their development towards DNS systems. Their most optimistic estimate is a merger density of up to $\SI{400}{\per\yr\per\cubic\giga\parsec}$, which corresponds to a merger rate of $\SI{0.88}{\per\yr}$. It is remarkable, that both these ab-initio estimates are clearly 
lower than the estimates resulting from LIGO or derived here. Nevertheless, the
difference is small enough that they are roughly 
compatible with the results from validation of the results by \cite{Pol_2020}
and our enhanced simulations using the generic profile, as can be seen in Fig.\ \ref{fig:DNSmergercompilation}.\\

\section{Summary \& Conclusion}

Improving on important
assumptions made in previous estimates of the Galactic DNS merger rate and the resulting
LIGO detection rate, we obtain results that are in good agreement with previous estimates
when considering the whole sample of DNS included in this study. This is in contrast to the
individual rate derived for PSR J1906+0746, which shows the expected large contribution
to the overall rates, and which increases using the measured beam shape in comparison to 
previous assumptions or generic profiles.  Applying generic profiles derived from PSR J1906+0746's
observation to the other DNSs makes, however, little overall difference in the total 
rate when combining the results.  
We  therefore conclude that the method of estimating the DNS merger and LIGO detection
rates via the study of the known Galactic radio pulsar DNS population is less prone
to systematic uncertainties than previously thought and argue that the 
derived estimates should be considered as robust. Consequently,
following the work presented here, we predict a detection of 3 to 9 DNS mergers per year by
Advanced LIGO (within 90\% confidence intervals). However, there are still a number of other
systematic uncertainties present in this and other methods, so that actual
results from the third operating run O3 of LIGO will provide significant insight in the
correctness of the different estimates and may allow us to address the various
caveats discussed earlier.

 In the case of PSR~J1906$+$0746, further measurements to improve the implemented intensity profile 
 can be made until the pulsar disappears from view, constraining our extrapolations due to 
 even better knowledge of the beam shape. The current measurements (Fig.\ \ref{fig:pulseprofile}) show significant differences in the shape of the main pulse and interpulse, imposing
 important constraints on pulsar emission models. Hence, future work on these aspects might result in interesting insights on the nature of pulsar beams and the beaming process itself.
 
Moreover, the differences between ab initio simulations and the simulations based on the
population of detected DNS systems could be resolved by including detected HMXB (high mass X-Ray binaries) systems or radio-quiet NS-star binaries into the detection based simulations, as they depict preliminary stages to DNS systems.

Finally, for a number of purposes, it would be tremendously helpful to expand the tracking of pulsars
showing relativistic spin precession, allowing us to have detailed studies such as for PSR~J1906+0746 also for
other DNS systems. This would not only help our understanding of the pulsar emission processes
and the general structure of pulsar emission beams, but it would also help to decrease the overall uncertainties of DNS merger rate predictions.

\section*{Acknowledgements}
We thank Nihan Pol for helpful discussions and provision of his simulation framework.
MK and GD are supported by the European Research Council for the ERC Synergy
Grant BlackHoleCam under contract no. 610058. We used the code to determine the SNR degradation factor of PSR J0508$+$3801 developed and written by \citet{2013MNRAS.432.1303B}.


\section*{Data Availability}
The data underlying this article will be shared on reasonable request to the corresponding author.



\bibliographystyle{mnras}
\bibliography{dnsmerger} 


\bsp	
\label{lastpage}
\end{document}